\documentclass[twocolumn,showpacs,preprintnumbers,amsmath,amssymb,aps,prb,groupedaddress]{revtex4-2}
\pdfoutput=1
\usepackage[pdftex]{graphicx}
\usepackage[english]{babel}
\usepackage{soul}
\usepackage{cleveref}
\usepackage{bbold}
\usepackage{mathtools}
\usepackage{multirow}
\usepackage{colortbl}
\definecolor{kugray5}{RGB}{224,224,224}
\usepackage[latin1]{inputenc}
\usepackage{diagbox}
\usepackage{mciteplus}

\usepackage{dcolumn}% Align table columns on decimal point\right] 
\usepackage{bm}% bold math

\usepackage{color}
\usepackage{amstext}
\usepackage{braket}
\usepackage{mathdots}
\usepackage{tikz}
\usepackage{physics}
\usepackage{enumitem}

\usetikzlibrary{matrix,decorations.pathreplacing}

\begin{document}

%%%%%%%%%%%%%%%%%%%%%%%%%%%%%%%%%%%%%%%%%%%%%%%%%%%%%%%%%%%%%%%%%%%%%%%%%%
%                               Title                                    %
%%%%%%%%%%%%%%%%%%%%%%%%%%%%%%%%%%%%%%%%%%%%%%%%%%%%%%%%%%%%%%%%%%%%%%%%%%

\title{Effects of Peierls phases in open linear chains}

\author{A. M. Marques}
\email{anselmomagalhaes@ua.pt}
\affiliation{Department of Physics $\&$ i3N, University of Aveiro, 3810-193 Aveiro, Portugal}

%\date{\today}

%%%%%%%%%%%%%%%%%%%%%%%%%%%%%%%%%%%%%%%%%%%%%%%%%%%%%%%%%%%%%%%%%%%%%%%%%%
%                              abstract                                  %
%%%%%%%%%%%%%%%%%%%%%%%%%%%%%%%%%%%%%%%%%%%%%%%%%%%%%%%%%%%%%%%%%%%%%%%%%%

\begin{abstract}
The introduction of Peierls phases in open tight-binding chains without closed paths in either real or synthetic dimensions is understood to be physically inconsequential, as one assumes they can always be gauged away.
Here, we show that this assumption does not necessarily hold for all systems in open chains, as closed paths may appear in the Fock space where these phases can lead to the creation of magnetic flux analogs with physical effects.
This idea is first illustrated in the quadratic Kitaev model, where different patterns for the Peierls phases are studied and their independent manipulation is seen to be able to drive the appearance of topological states and Majorana flat bands.
We then consider a system with quartic interactions, namely an extended Bose-Hubbard (EBH) open chain with a finite Peierls phase associated with the hopping terms.
Focusing on the strong interactions limit of the two-body sector, we show the decisive influence of these phases on the two-band spectrum of the higher energy subspace, which behaves as an effective sawtooth chain with magnetic flux at each plaquette.
In particular, both the width of the energy gap between the two bands and the position of the in-gap edge states can be controlled by the Peierls phase.
Finally, by translating this two-particle one-dimensional (1D) system into a single-particle two-dimensional (2D) one, and subsequently mapping it onto an equivalent electrical LC circuit, we identify a parameter set for which in-gap states are only present for certain finite phase values.
Tuning the circuit to these parameters generates the corresponding boundary voltage states, providing an experimentally detectable signature of the effects induced by manipulating Peierls phases in a model built on an open linear chain. 
\end{abstract}

%\pacs{74.25.Dw,74.25.Bt}

\maketitle
% SECTION
%%%%%%%%%%%%%%%%%%%%%%%%%%%%%%%%%%%%%%%%%%%%%%%%%%%%%%%%%%%%%%%%%%%%%
%%%%%%%%%%%%%%%%%%%%%%%%%%%%%%%%%%%%%%%%%%%%%%%%%%%%%%%%%%%%%%%%%%%%%
%%%%%%%%%%%%%%%%%%%%%%%%%%%%%%%%%%%%%%%%%%%%%%%%%%%%%%%%%%%%%%%%%%%%%
\section{Introduction}
\label{sec:intro}
Typically, in single-particle models the global Peierls phase picked up along a direction obeying open boundary conditions (OBC) can be gauged away, even though this may not be true locally due to the presence of closed paths (e.g., in quasi-1D models such as an open diamond chain, where a finite magnetic flux may thread the plaquettes locally while vanishing globally \cite{Mukherjee2018,Pelegri2019a,Pelegri2019b}).
Different behavior occurs, on the other hand, in many-body systems where, even for OBC in real space, closed paths involving specific hoppings may appear when the adjacency graph of the many-body Fock space is considered \cite{Santos2019,Balas2020,Santos2020}.
Including Peierls phases at the single-particle hoppings can translate into the appearance of magnetic flux analogs threading the Fock space along different directions that \textit{cannot} be gauged away, as they have physical consequences.
In fact, this effect can be seen already at the level of quadratic systems with simple analytical solutions, as we illustrate here by revisiting the Kitaev model in its original formulation \cite{Kitaev2001}, known to host non-local Majorana fermions under OBC in the topological phase \cite{Leijnse2012}, and discuss the consequences of introducing Peierls phases in the hopping parameters.

We then move to interacting systems, studying an EBH open uniform chain, which includes Hubbard and nearest-neighbor (NN) interactions \cite{Mazzarella2006,Dutta2015}, as well as superconducting-type pair hopping terms.
For strong interactions and two particles, the subspace of high-energy states can be modeled by an effective sawtooth chain, with the Peierls phases translating as a magnetic flux piercing each plaquette, 
which are therefore able to fundamentally change the energy profile of this subspace, thus opening an avenue for the experimental probing of their effects.
This should be contrasted with (i) the effect of these phases in \textit{periodic} interacting systems \cite{Mateos2023}, since in this case they just reflect the presence of an external magnetic flux, (ii) with many-body systems \cite{Nicolau2023a} where Peierls phase accumulation derived from perturbation theory generates subspaces with effective flux per plaquette, since this flux is related with the synthetic flux already present in each plaquette in the single-particle picture, and (iii) with the presence of persistent currents in small normal-metal mesoscopic rings \cite{Buttiker1983,Riedel1989}, that is, with a circumference smaller than the dephasing and thermal lengths, when pierced by an Aharonov-Bohm magnetic field, that have been experimentally detected \cite{Levy1990,Mailly1993}, which is again a consequence both of the periodic nature of the physical system and of the application of a real flux (even if the currents persist once the flux is turned off).

Several works on the two-body physics in 1D interacting systems have already shown how these models can be exactly mapped onto equivalent single-particle 2D systems \cite{Valiente2008,Longhi2011,Krimer2011,Corrielli2013,Mukherjee2016,DiLiberto2016,Gorlach2017a,Gorlach2018}, where the movement of each particle of the original system generates one of the orthogonal coordinates of the mapped one, and the interactions are translated as local potentials.
In general, the problem of $M$ interacting particles in a 1D linear chain can be mapped onto a single-particle living in an $M$-dimensional lattice \cite{Cheng2021}.
For few-body systems, where the resulting mapped dimensionality is not too high, the main advantages of this technique are two-fold: by getting rid of the interactions (i) the problem becomes much more amenable to an analytical treatment, and (ii) the effects of the interactions in the original model are easier to simulate experimentally in artificial lattices, such as electrical circuits \cite{Olekhno2020}, designed to realize the non-interacting mapped model.
As such, we will also take advantage of this technique to map the interacting EBH chain onto a non-interacting square lattice, further showing how the latter can be implemented using LC circuits.
The appearance of in-gap corner voltage states when OBC are applied to the circuit can be understood as a direct consequence of the finite Peierls phases included in the original EBH chain, and therefore the physical effects of the phases can be experimentally tested and validated through the detection of these corner modes.

The rest of the paper is organized as follows.
In Sec.~\ref{sec:kitaev}, we consider the open Kitaev chain with different patterns for the Peierls phases at the hopping parameters and analyze how they affect the energy spectrum and the topological properties of the model.
In Sec.~\ref{sec:ebhchain}, we introduce a modified version of the open EBH chain, and pay particular attention to the subspace of high-energy states that appears in the strong interactions limit.
Up to first-order in perturbation theory, this subspace is shown to model a sawtooth chain, with the original Peierls phases translating as a magnetic flux at each plaquette, one whose variation changes the energy spectrum in a decisive way.
In Sec.~\ref{sec:lccircuit}, after outlining the basic theory behind topolectrical circuits \cite{Lee2018}, we show how the EBH chain studied in the previous section can be described by an LC circuit forming a 2D lattice.
Based on this mapping and corresponding numerical simulations, we propose an experimental scheme to detect the corner voltage states that arise in the admittance gap as a consequence, in the original EBH chain, of the finite Peierls phases.
Finally, in Sec.~\ref{sec:conclusions} we present our concluding remarks.

% SECTION
%%%%%%%%%%%%%%%%%%%%%%%%%%%%%%%%%%%%%%%%%%%%%%%%%%%%%%%%%%%%%%%%%%%%%
%%%%%%%%%%%%%%%%%%%%%%%%%%%%%%%%%%%%%%%%%%%%%%%%%%%%%%%%%%%%%%%%%%%%%
%%%%%%%%%%%%%%%%%%%%%%%%%%%%%%%%%%%%%%%%%%%%%%%%%%%%%%%%%%%%%%%%%%%%%
\section{Peierls phases in the Kitaev chain}
\label{sec:kitaev}
The Hamiltonian of the Kitaev chain models a 1D chain placed on the surface of a p-wave superconductor, which induces spin triplet pair tunnelings to the chain by proximity effect, and reads as
\begin{eqnarray}
	H&=&\sum\limits_{j=1}^{N-1}\Big[-te^{i\phi_t^j}c^\dagger_{j+1}c_j+\Delta e^{i\phi_\Delta^j}c^\dagger_{j+1}c^\dagger_j+H.c.\Big] \nonumber
	\\
	&+&\mu\sum\limits_{j=1}^Nc^\dagger_j c_j ,
	\label{eq:hamilt}
\end{eqnarray}
where $N$ is the number of sites in the chain with OBC, $c_j$ is the spinless fermionic destruction operator acting on site $j$, $t$ and $\Delta$ are, respectively, the magnitudes of the hopping and superconducting pairing parameters, $\mu$ is the chemical potential, and we include local Peierls ($\phi_t^j$) and superconducting ($\phi_\Delta^j$) phases as free parameters.
Throughout this paper we set $\mu=0$ and $t=\Delta>0$.

\begin{figure}[ht]
	\begin{centering}
		\includegraphics[width=0.45\textwidth]{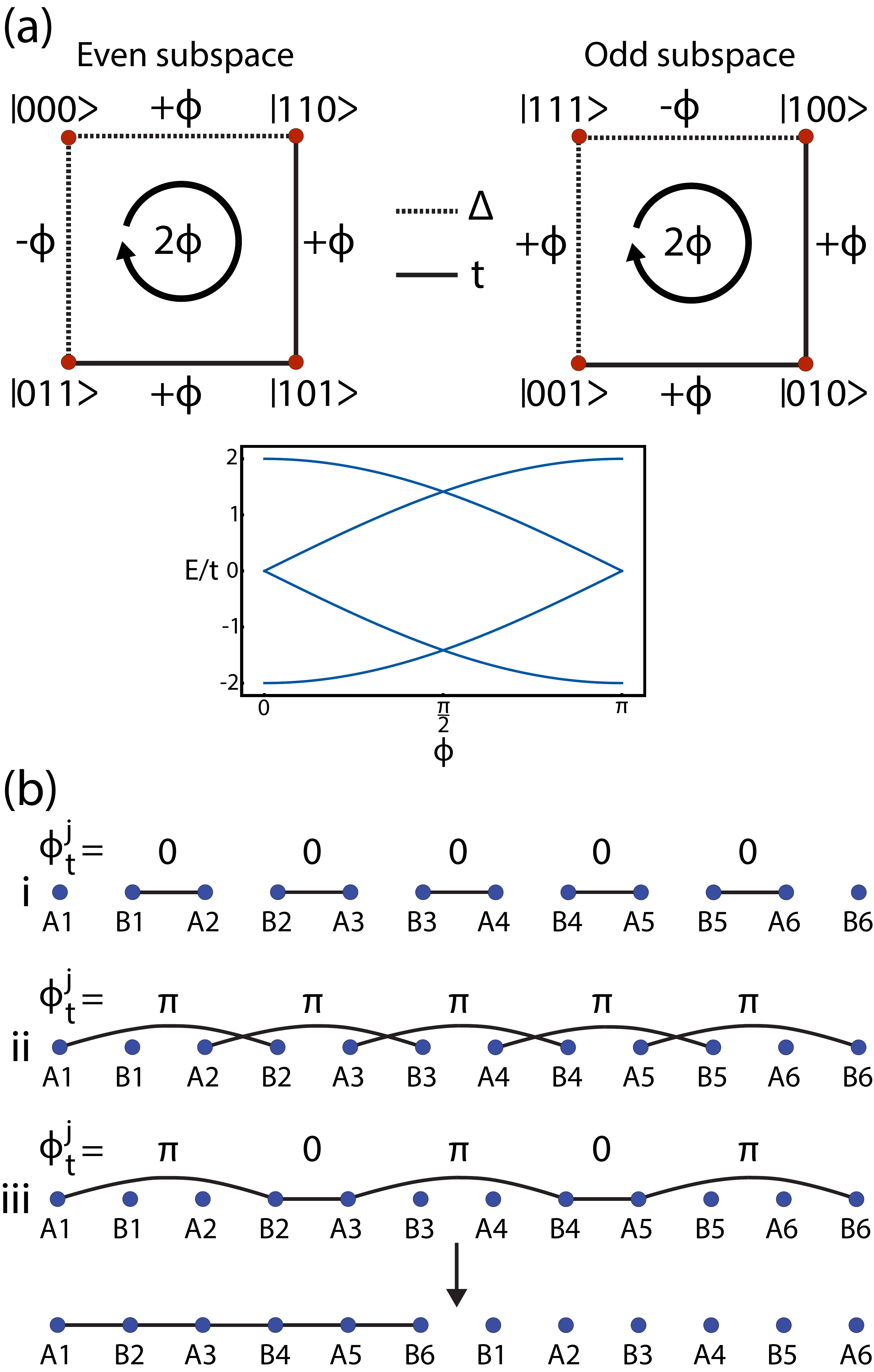} 
		\par\end{centering}
	\caption{(a) Graph of the even and odd Fock subspaces of the Kitaev chain defined in (\ref{eq:hamilt}) for N=3 and $\phi_\Delta^j=\phi_t^j=\phi$. The respective energy spectrum as a function of $\phi$ is shown below, with each curve two-fold degenerate. (b) Depiction of the Kitaev chain with $N=6$ in the basis of Majorana operators for i) $\phi_t^j=0$, ii) $\phi_t^j=\pi$, and iii) $\phi_t^j=\pi j \mod 2\pi$.}
	\label{fig:3sites}
\end{figure}

To illustrate the non-trivial effect of including Peierls phases in the system, let us consider an open chain with $N=3$ sites and constant phases $\phi_t^j=\phi_\Delta^j=\phi$ \cite{Degottardi2013}. 
When one goes from the real-space basis to the many-body basis where the states are written as, e.g., $c^\dagger_2c^\dagger_3\ket{000}=\ket{011}$, with $\ket{000}$ the state of the empty chain, the real open chain becomes a closed graph \cite{Roy2019,Roy2020} for the two decoupled and degenerate even and odd Fock subspaces, as shown in Fig.~\ref{fig:3sites}(a).
Around these loops, the superconducting phases are seen to cancel out while the Peierls phases accumulate, creating a $2\phi$ flux threading each subspace.
The energy spectrum obtained from exact diagonalization further corroborates the non-negligible effects of introducing Peierls phases in an open Kitaev chain.
Note that $N=3$ sites is the minimum number of sites for which loop structures appear in the even and odd subspaces.
For example, for the recently realized two-site minimal Kitaev chain \cite{Dvir2023} in a system of coupled quantum dots \cite{Sau2012}, both the Peierls and superconducting phases can be gauged away \cite{Samuelson2024}, since the graph of both subspaces yields a dimer: one connecting the $\ket{00}$ and $\ket{11}$ ($\ket{01}$ and $\ket{10}$) states by $\Delta$ ($t$) for the even (odd) subspace.

The Hamiltonian in (\ref{eq:hamilt}) can be written in terms of Majorana operators, defined as $\gamma_{j}^A=c^\dagger_j+c_j$ and $\gamma_{j}^B=i(c^\dagger_j-c_j)$, with $\gamma_{j}^\eta\equiv\gamma_{j}^{\eta^\dagger}$, and obeying the Clifford algebra $\{\gamma_{i}^\eta,\gamma_{j}^\mu\}=2 \delta_{\eta,\mu}\delta_{i,j}$, with $\delta_{i,j}$ the Kronecker delta.
Setting $\phi_\Delta^j=0$, one arrives at
\begin{equation}
H=it\sum\limits_j\gamma^\eta_{j}\gamma_{j+1}^\mu,
\end{equation}
with $(\eta,\mu)=(B,A)$ for $\phi_t^j=0$ [see Fig.~\ref{fig:3sites}(b)i] and $(\eta,\mu)=(A,B)$ for $\phi_t^j=\pi$ [see Fig.~\ref{fig:3sites}(b)ii],
reflecting the fact that a global $\pi$ change in the Peierls phases can be absorbed in a global change in the indexation ($A\leftrightarrow B$). 
In both cases, the Majorana operators of the Kitaev chain are known to form dimers at the bulk, leaving two decoupled Majorana zero modes (MZMs) at each end, with both combining to create a Majorana fermion.
When one considers an alternating pattern for the Peierls phases of the form $\pi 0\pi 0\dots$, given explicitly by $\phi_t^j=\pi j \mod 2\pi$, the Hamiltonian becomes
\begin{equation}
H=it\Big[\sum\limits_{j=1}^{N_\pi}\gamma_{2j-1}^A\gamma_{2j}^B +\sum\limits_{j=1}^{N_0}\gamma_{2j}^B\gamma_{2j+1}^A\Big],
\label{eq:hamiltmajalt}
\end{equation}
where $N_{\pi(0)}$ is the number of hoppings with a $\pi(0)$ Peierls phase, with $N_\pi+N_0=N-1$. 
As illustrated in Fig.~\ref{fig:3sites}(b)iii, in this case half ($N$) of the Majorana operators become decoupled, while the other $N$ form the linear chain expressed by (\ref{eq:hamiltmajalt}), with a Majorana dispersion relation $E_k=-t\cos k$ and corresponding diagonalized Majorana operators $\gamma_k=\sqrt{\frac{2}{N+1}}\sum_{j=1}^N(-i)^j\sin(kj)\gamma_j^\eta$, where $k_n=\frac{n\pi}{N+1}$ with $n=1,2,\dots,N$ from the OBC, $\eta=A(B)$ for $j$ odd (even). The modulated phase factor in the sum comes from the fact that the $i$ factor in (\ref{eq:hamiltmajalt}) cannot be dropped due to the $\gamma_j\equiv\gamma^\dagger_j$ property.
When $N$ is even (implying an integer number of unit cells, as we will see), the remaining $N$ decoupled operators can form $N/2$ pairs of degenerate localized Majorana fermions, that is, they originate zero energy flats bands \footnote{When N is odd, $\frac{N+1}{2}$ Majorana fermions are still formed since one gets an extra zero energy state coming from the dispersive band for $k=\frac{\pi}{2}$, given by $\tilde{\gamma}^A=\gamma_{\frac{\pi}{2}}=\sqrt{\frac{2}{N+1}}\sum_{j=1}^{\frac{N+1}{2}}\gamma_{2j-1}^A$, which corresponds to a highly delocalized Majorana operator with equal weight on all the $\gamma^A$ of the dispersive chain}.
From the relation $\gamma_{k_n}^\dagger=\gamma_{k_{N+1-n}}$, fermionic modes can be constructed from the dispersive band by defining the operators $\tilde{c}_{k_n}=\frac{1}{\sqrt{2}}\gamma_{k_n}$ and $\tilde{c}_{k_n}^\dagger=\frac{1}{\sqrt{2}}\gamma_{k_{N+1-n}}$, with $n=1,2,\dots,N/2$ and $\{\tilde{c}_k,\tilde{c}_{k^\prime}^\dagger\}=\delta_{kk^\prime}$, leading to
\begin{equation}
H=\sum\limits_{n=1}^{N}E_{k_n}\gamma_{k_{N+1-n}}\gamma_{k_n}=\sum\limits_{n=1}^{N/2}4E_{k_n}\Big(\tilde{c}_{k_n}^\dagger\tilde{c}_{k_n}-\frac{1}{2}\Big),
\label{eq:diaghamilt}
\end{equation}
where $E_{k_n}=-E_{k_{N+1-n}}$ was used and $k_n\in(0,\frac{\pi}{2})$ in the Hilbert space of halved dimension of the full fermionic modes.

\begin{figure}[ht]
	\begin{centering}
		\includegraphics[width=0.48\textwidth]{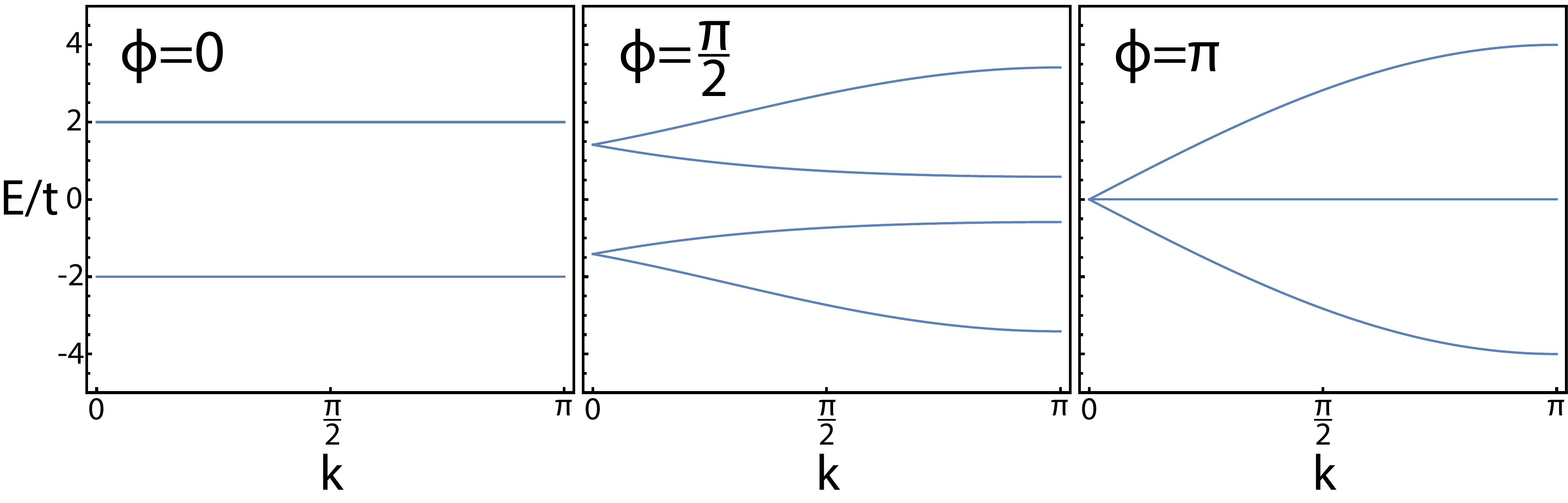} 
		\par\end{centering}
	\caption{Energy spectrum of the Kitaev chain with an alternating $\phi 0\phi 0\dots$ pattern for the Peierls phases at the hopping parameters for $\phi=0,\frac{\pi}{2},\pi$. Each flat band is two-fold degenerate.}
	\label{fig:espectrum_twosites}
\end{figure}

The appearance of Majorana flat bands can be checked by diagonalization of (\ref{eq:hamilt}) under periodic boundary conditions (PBC). 
A general $\phi 0\phi 0\dots$ pattern for the Peierls phases, given by $\phi_t^j=\phi j \mod 2\phi$, implies a two-sites unit cell, with the sites labeled $\alpha$ and $\beta$, where PBC are satisfied by setting $c^\dagger_{\alpha(\beta)1}\equiv c^\dagger_{\alpha(\beta)N_{uc}+1}$, with $N_{uc}$ the number of unit cells. 
Upon Fourier transforming into $k$-space with $c_{\alpha(\beta)j}=\frac{1}{\sqrt{N_{uc}}}\sum_k e^{ikj}c_{\alpha(\beta)k}$ and lattice spacing $a\equiv1$, and defining $\Psi_k=(c_{\alpha k}\  c_{\beta k}\ c^\dagger_{\alpha -k}\ c^\dagger_{\beta -k})^T$, the Hamiltonian with $\phi_{\Delta}=0$ everywhere becomes
\begin{eqnarray}
H&=&\sum\limits_{k>0} \Psi_k^\dagger\mathcal{H}_k\Psi_k,
\label{eq:hamiltk}
\\
\mathcal{H}_k&=&
\begin{pmatrix}
\Sigma_k&\Gamma_k
\\
\Gamma_k^\dagger&-\Sigma_{-k}^T
\end{pmatrix},
\end{eqnarray}
where $\Sigma_k=-\mu\sigma_0-t(\cos\phi+\cos k)\sigma_x-t(\sin\phi+\sin k)\sigma_y$, $\Gamma_k=i\sin k\sigma_x+i(1-\cos k)\sigma_y$, $\sigma_\nu$ $(\nu=x,y,z)$ is the $\nu$ Pauli matrix, and $\sigma_0$ is the $2\times 2$ identity matrix.
Note that time-reversal symmetry (TRS), defined as $\mathcal{K}\mathcal{H}_k\mathcal{K}^{-1}=\mathcal{H}_{-k}$, where $\mathcal{K}$ is the complex conjugation operator obeying $\mathcal{K}\mathcal{K}^{-1}=1$, is broken for $\phi\neq 0,\pi$.
Similarly, inversion symmetry, defined as $P\mathcal{H}_k P^{-1}=\mathcal{H}_{-k}$, where $P=\sigma_z\otimes\sigma_x$, is only recovered for $\phi=0,\pi$.
Diagonalization of this Hamiltonian with the parameters considered in this study ($\mu=0$ and $t=\Delta>0$) yields four energy bands,
\begin{widetext}
\begin{equation}
	E^{\pm\pm}(k,\phi)=\pm t\sqrt{4+2(\cos\phi -1)\cos k\pm 4\sqrt{\Big(3+(\cos\phi -1)\cos k+\cos\phi\Big)\sin(\frac{k}{2})^2\sin(\frac{\phi}{2})^2}}.
\end{equation}
\end{widetext}
The energy spectra for three different values of $\phi$ are shown in Fig.~\ref{fig:espectrum_twosites}.
For the $\phi=0$ case at the left panel the primitive cell is given by a single site, so that our two-sites unit cell represents, in relation to that, a folding of the reduced Brillouin zone, defined as $k\in(0,\pi]$, and a concomitant doubling of the number of bands. In the Majorana basis, this model corresponds to the one shown in Fig.~\ref{fig:3sites}(b)i under OBC, whose uncoupled MZMs at both ends are known to stem from the non-trivial topology of this model.
As $\phi$ increases all four bands become dispersive and the degeneracies are lifted everywhere, since $k=0$ is outside the reduced Brillouin zone [see middle panel of Fig.~\ref{fig:espectrum_twosites} for $\phi=\frac{\pi}{2}$].
When $\phi=\pi$ is reached, the two middle bands coalesce into a doubly degenerate Majorana flat band [see right panel of Fig.~\ref{fig:espectrum_twosites}].
Simultaneously, the zero-energy band gap closes, marking the critical point of the topological transition.

Additional insight into the model with a $\phi 0\phi 0\dots$ pattern for the Peierls phases and $\phi_\Delta^j=0$ can be obtained by changing the gauge to $c_{\alpha_j}\to e^{-i\phi j}c_{\alpha_j}$, with $\alpha_j=2j,2j+1$ and $j=1,2,\dots,\frac{N}{2}-1$, $N$ even and boundaries $c_1\to c_1$ and $c_N\to e^{-i\phi\frac{N}{2}}c_N$.
Under this gauge transformation, the Peierls phases vanish everywhere, $\phi_t^j=0$, and the superconducting phases are now given by $\phi_\Delta^j=\phi j$, that is, the model picks up a constant superconducting phase gradient (SPG), with $\Delta c^\dagger_{j+1}c^\dagger_j\to \Delta e^{i\phi j}c^\dagger_{j+1}c^\dagger_j$ in (\ref{eq:hamilt}).
Recent studies have addressed similar quadratic models in the presence of an SPG induced by a bulk supercurent across the superconducting substrate \cite{Romito2012,Sticlet2013,Liu2013,Dmytruk2019,Melo2019}.
In particular, Mahyaeh and Ardonne \cite{Mahyaeh2018} considered spinless fermions and studied the interplay between an SPG and complex long-range hoppings (which introduce closed paths that can be threaded by finite fluxes already at the level of a single-particle system \cite{Niu2012,Wang2016,Nehra2020,Soori2024} where, in contrast, the effect of the Peierls phases we introduce here only manifests itself in the Fock space).
The authors considered a different gauge transformation that converts a constant SPG into constant Peierls and superconducting phases across the chain (or into Peierls and spin-orbit phases in similar spinfull models \cite{Romito2012,Chevallier2013}) and presented a thorough analysis of the topological phases supported by their model.
However, it appears that a study on the independent effect of the Peierls phases in the absence of real-space closed paths (or internal-space closed paths such as those created by long-range hoppings \cite{Viyuela2016,Amin2019}), has not been performed thus far.

By taking advantage of the formal equivalence between a staggered $\phi 0\phi 0\dots$ pattern for the Peierls phases and a constant SPG in a linear Kitaev chain, an alternative approach to simulate the effects of the latter is provided.
After the gauge transformation introduced above, the same staggered pattern in the Peierls phases appears on the superconducting phases for $\phi=\pi$, therefore leading to the results depicted in Fig.~\ref{fig:3sites}(b)iii and at the right panel of Fig.~\ref{fig:espectrum_twosites}.
It was shown that in this case the model is topologically trivial in the entire phase space \cite{Mahyaeh2018}. 
For completeness sake, we also note that the case of constant Peierls phases, $\phi_t^j=\phi$, can be translated into a constant SPG with $\phi_\Delta^j=-2\phi j$ through the gauge transformation $c_j\to e^{-i\phi(j-\frac{1}{2})}c_j$.

% SUBSECTION
%%%%%%%%%%%%%%%%%%%%%%%%%%%%%%%%%%%%%%%%%%%%%%%%%%%%%%%%%%%%%%%%%%%%%
%%%%%%%%%%%%%%%%%%%%%%%%%%%%%%%%%%%%%%%%%%%%%%%%%%%%%%%%%%%%%%%%%%%%%
%%%%%%%%%%%%%%%%%%%%%%%%%%%%%%%%%%%%%%%%%%%%%%%%%%%%%%%%%%%%%%%%%%%%%
\subsection{Flat bands and topology}
Let us consider now the case of a three-sites unit cell with $\phi_\Delta^j=0$ for all $j$ and a $00\phi 00\phi\dots$ pattern for the Peierls phases, given explicitly by $\phi_t^j=\phi(1+2\cos(\frac{2\pi}{3}j))/3$. Defining $\Psi_k=(c_{\alpha k}\  c_{\beta k}\ c_{\gamma k}\ c^\dagger_{\alpha -k}\ c^\dagger_{\beta -k}\ c^\dagger_{\gamma -k})^T$, the Bloch Hamiltonian $\mathcal{H}_k$ in (\ref{eq:hamiltk}) becomes now
\begin{eqnarray}
\mathcal{H}_k&=&
-t
\begin{pmatrix}
\Sigma_k&\Gamma_k
\\
\Gamma_k^\dagger&-\Sigma_{-k}^T
\end{pmatrix},
\label{eq:hbulk3}
\\
\Sigma_k&=&\lambda_1+\lambda_6+\cos(k-\phi)\lambda_4+\sin(k-\phi)\lambda_5,
\\
\Gamma_k&=&-i(\lambda_2+\lambda_7)+i\cos k\lambda_5-i\sin k\lambda_4,
\end{eqnarray}
where $\lambda_i$ is the $i^{th}$ $3\times 3$ Gell-Mann matrix,
\begin{eqnarray}
		\lambda_1&=&
		\begin{pmatrix}
			0&1&0
			\\
			1&0&0
			\\
			0&0&0
		\end{pmatrix},\ \ 
		\lambda_2=
		\begin{pmatrix}
			0&-i&0
			\\
			i&0&0
			\\
			0&0&0
		\end{pmatrix}, \nonumber
		\\
		\lambda_4&=&
		\begin{pmatrix}
			0&0&1
			\\
			0&0&0
			\\
			1&0&0
		\end{pmatrix},\ \ 
		\lambda_5=
		\begin{pmatrix}
			0&0&-i
			\\
			0&0&0
			\\
			i&0&0
		\end{pmatrix}, \nonumber
		\\
		\lambda_6&=&
		\begin{pmatrix}
			0&0&0
			\\
			0&0&1
			\\
			0&1&0
		\end{pmatrix},\ \ 
		\lambda_7=
		\begin{pmatrix}
			0&0&0
			\\
			0&0&-i
			\\
			0&i&0
		\end{pmatrix},
\end{eqnarray}
and $t=\Delta>0$, $\mu=0$ and $\phi_\Delta^j=0$ is already assumed.
\begin{figure}[h]
	\begin{centering}
		\includegraphics[width=0.48\textwidth]{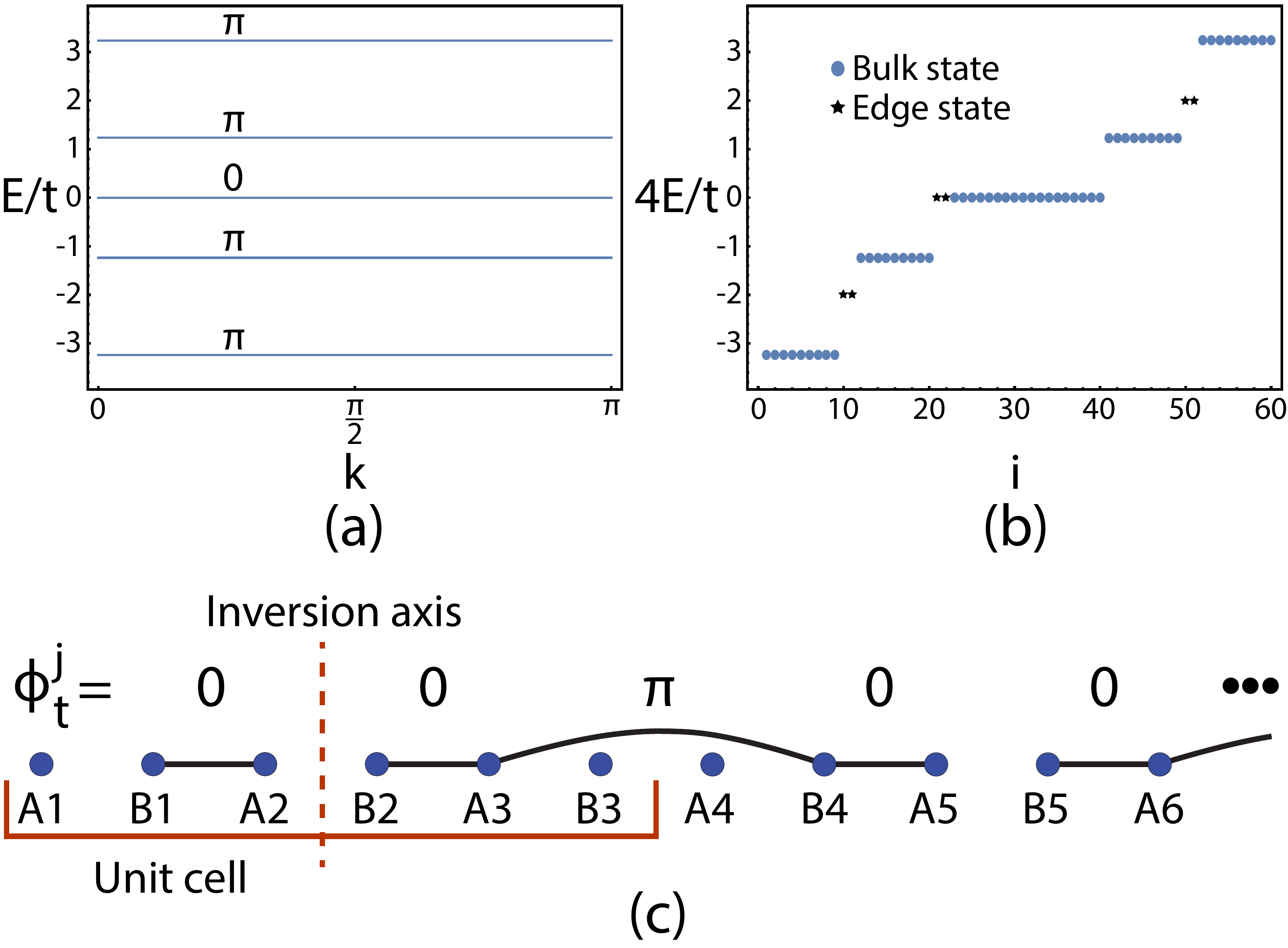} 
		\par\end{centering}
	\caption{(a) Energy spectrum of the Kitaev chain with an alternating $00\pi 00\pi\dots$ pattern for the Peierls phases at the hopping parameters. The zero-energy flat band is two-fold degenerate. Values $\alpha=0,\pi$ indicate the Zak phase of the respective band (cumulative for the degenerate zero-energy bands).
	(b) Majorana spectrum for the model in (a), multiplied by 4 for comparison, with $N=30$ and OBC in the Majorana basis.
    (c) Partial depiction, in the basis of Majorana operators, of the model with the Majorana spectrum of (b).}
	\label{fig:espectrum_threesites}
\end{figure}
Diagonalization of (\ref{eq:hbulk3}) yields six flat bands for all $\phi$. In particular,
when $\phi=\pi$ we find $E_1=-E_6=-(\sqrt{5}+1)t$, $E_2=-E_5=-(\sqrt{5}-1)t$ and two Majorana flat bands $E_3=E_4=0$, as shown in the energy spectrum of Fig.~\ref{fig:espectrum_threesites}(a).
Topological information can be extracted from the computation of the Zak phase for each band,
\begin{equation}
\mathcal{Z}_j=i\int_{-\pi}^{\pi}\bra{u_j(k)}\frac{d}{dk}\ket{u_j(k)},
\label{eq:zak}
\end{equation}
where $\ket{u_j(k)}$ is the eigenvector of band $j$ and the integral goes over the full Brillouin Zone. In the case of the degenerate Majorana bands in Fig.~\ref{fig:espectrum_threesites}(a), a cumulative Zak phase for both bands is considered.
The topological nature of each gap is given by the sum of the Zak phases of all occupied bands below it modulo 2$\pi$. 
This way, the top and bottom energy gaps are identified as the only topologically non-trivial gaps.
In order to check if topological edge states appear in these gaps we turn to the basis of Majorana operators under OBC, with a six-sites unit cell as depicted in Fig.~\ref{fig:espectrum_threesites}(c), where the Hamiltonian is written as
\begin{eqnarray}
H&=&it\sum\limits_{j=1}^{N_{uc}}\big[\gamma_{3j-2}^B\gamma_{3j-1}^A+\gamma_{3j-1}^B\gamma_{3j}^A\big] \nonumber
\\
&+&it\sum\limits_{j=1}^{N_{uc}-1}\gamma_{3j}^A\gamma_{3(j+1)-2}^B.
\end{eqnarray}
Upon diagonalization, the energy spectrum of this Hamiltonian for $N_{uc}=10$ is shown in Fig.~\ref{fig:espectrum_threesites}(b).
Notice that the energies are multiplied by a factor of 4, which was included for comparison with Fig.~\ref{fig:espectrum_threesites}(a).
This is required since the diagonalized $\gamma$ operators constitute ``half'' fermion states with quarter energies in relation to the full fermion states that are obtained by combining $\gamma$ states with symmetric energies, as in (\ref{eq:diaghamilt}).
States of the four flat bands with finite energy are localized along the decoupled clusters at the bulk of the chain [e.g., see the four-sites cluster formed by B2-A3-B4-A5 in Fig.~{\ref{fig:espectrum_threesites}(c)}].
States in the two degenerate Majorana flat bands come from the two free $\gamma$ operators appearing below each ``$\pi$-bridge'' at the bulk [see decoupled sites B3 and A4 in Fig.~\ref{fig:espectrum_threesites}(c)].
In agreement with the Zak phases of the bands in Fig.~\ref{fig:espectrum_threesites}(a), pairs of topological edge $\gamma$ states are found lying within both the top and bottom energy gaps in Fig.~\ref{fig:espectrum_threesites}(b).
These are states localized at the dimers appearing only at the edge unit cells [see dimer of sites B1-A2 in Fig.~\ref{fig:espectrum_threesites}(c)].
Lastly, there are two extra MZMs at the decoupled A1 and B$_{3N_{uc}}$ sites which can be combined to form the usual Majorana fermion.
However, since these states are not in-gap states, being instead degenerate with the bulk MZMs of the zero-energy flat bands, they are not topologically protected.

There are two other choices for the unit cell, corresponding to the $0\pi 00\pi 0\dots$ and $\pi 0 0\pi 00\dots$ patterns for the Peierls phases.
Even though the system is always inversion-symmetric, for these other choices of unit cell the inversion-axes in the Majorana basis do not cross the center of the unit cell which, as shown elsewhere \cite{Marques2019}, entails non $\pi$-quantized Zak phases for the bands using (\ref{eq:zak}) \cite{Marques2018,Pelegri2019a,Pelegri2019b,Madail2019,Kremer2020}. 
Contrary to the model with a two-sites unit cell studied before, in this case there is no gauge transformation capable of fully absorbing the Peierls phases into a constant SPG.
As such, and apart from less realistic fine tuned non-constant SPGs, there is no equivalent modulation of the superconducting phases capable of recreating the energy spectrum of Kitaev chains with more complex patterns for the Peierls phases.

%\textbf{Topological characterization?} For the $\phi=0$ case at the left panel, our model recovers chiral symmetry, defined as $\hat{C}H_k\hat{C}^{\dagger}=-H_k$, where $\hat{C}=\sigma_x\otimes\sigma_0$.
%Changing to the basis where $\hat{C}=\mbox{diag}(1,1,-1,-1)$, the Hamitonian becomes block anti-diagonal,
%\begin{equation}
%	\hat{T}H_k\hat{T}^\dagger=
%	\begin{pmatrix}
%	0&h_k
%	\\
%	h_k^\dagger&0
%	\end{pmatrix},
%\end{equation}
%where $\hat{T}$ is the matrix whose columns correspond to the eigenvectors of $\hat{C}$ and $h_k=$

Finally, one should note that a ``$\pi$-bridge'', defined here as a $\pi$ Peierls phase sandwiched between 0 Peierls phases at the adjacent hopping parameters [see Fig.~\ref{fig:3sites}(b)iii], is formally equivalent to a double local superconducting $\pi$-junction.
To see this, let us suppose that an open Kitaev chain with N sites has zero Peierls and superconducting phases everywhere, with the exception of $\phi_t^j=\pi$, where $j$ is the index of some bulk hopping term. 
Then, by applying the following gauge transformation, $c_{l}\to -c_{l}$, for all $l>j$, results in $\phi_t^j=0$ and $\phi_\Delta^j=\pi$, that is, a $\pi$ superconducting phase shift occurs both at the left (between $j-1$ and $j$ sites) and at the right (between $j$ and $j+1$ sites).
Given that an MZM is trapped at a $\pi$-junction, such as the spurious MZMs that may appear at the crossings in nanowire junctions \cite{Alicea2011,Stanescu2018}, each ``$\pi$-bridge'' traps two MZMs, as exemplified in Fig.~\ref{fig:3sites}(b)iii.
More generally, the ability to control the signs of the hopping parameters can be seen as an alternative way to study the effects of local double $\pi$-junctions, which can be pinpointed at specific sites of the chain.

% SECTION
%%%%%%%%%%%%%%%%%%%%%%%%%%%%%%%%%%%%%%%%%%%%%%%%%%%%%%%%%%%%%%%%%%%%%
%%%%%%%%%%%%%%%%%%%%%%%%%%%%%%%%%%%%%%%%%%%%%%%%%%%%%%%%%%%%%%%%%%%%%
%%%%%%%%%%%%%%%%%%%%%%%%%%%%%%%%%%%%%%%%%%%%%%%%%%%%%%%%%%%%%%%%%%%%%
\section{Extended Bose-Hubbard chain}
\label{sec:ebhchain}

So far we have been addressing the effects of introducing Peierls phases at the hopping term of the Kitaev chain which, even though it allows for a representation in the Fock space where the effect of these phases becomes evident, remains a model with a quadratic Hamiltonian.
As such, and strictly speaking, it is not necessary to leave the single-particle picture in order to solve the model.
In this section, on the other hand, we introduce a simple toy model of an interacting many-body linear chain and show how the inclusion of Peierls phases at the hopping terms can lead to nontrivial phenomena.
We will further focus on the two-body sector for simplicity.
%%%%%%%%%%%%%%%%%%%%%%%%%%%%%%%%%%%%%%%%%%%%%%%%%%%%%%%%%%%%%%%%%
%%%%%%%%%%%%%%%%% Figure %%%%%%%%%%%%%%%%%%%%%%%%%%%%%%%%%%%%%%%%
%%%%%%%%%%%%%%%%%%%%%%%%%%%%%%%%%%%%%%%%%%%%%%%%%%%%%%%%%%%%%%%%%
\begin{figure}[ht]
	\begin{centering}
		\includegraphics[width=0.48\textwidth]{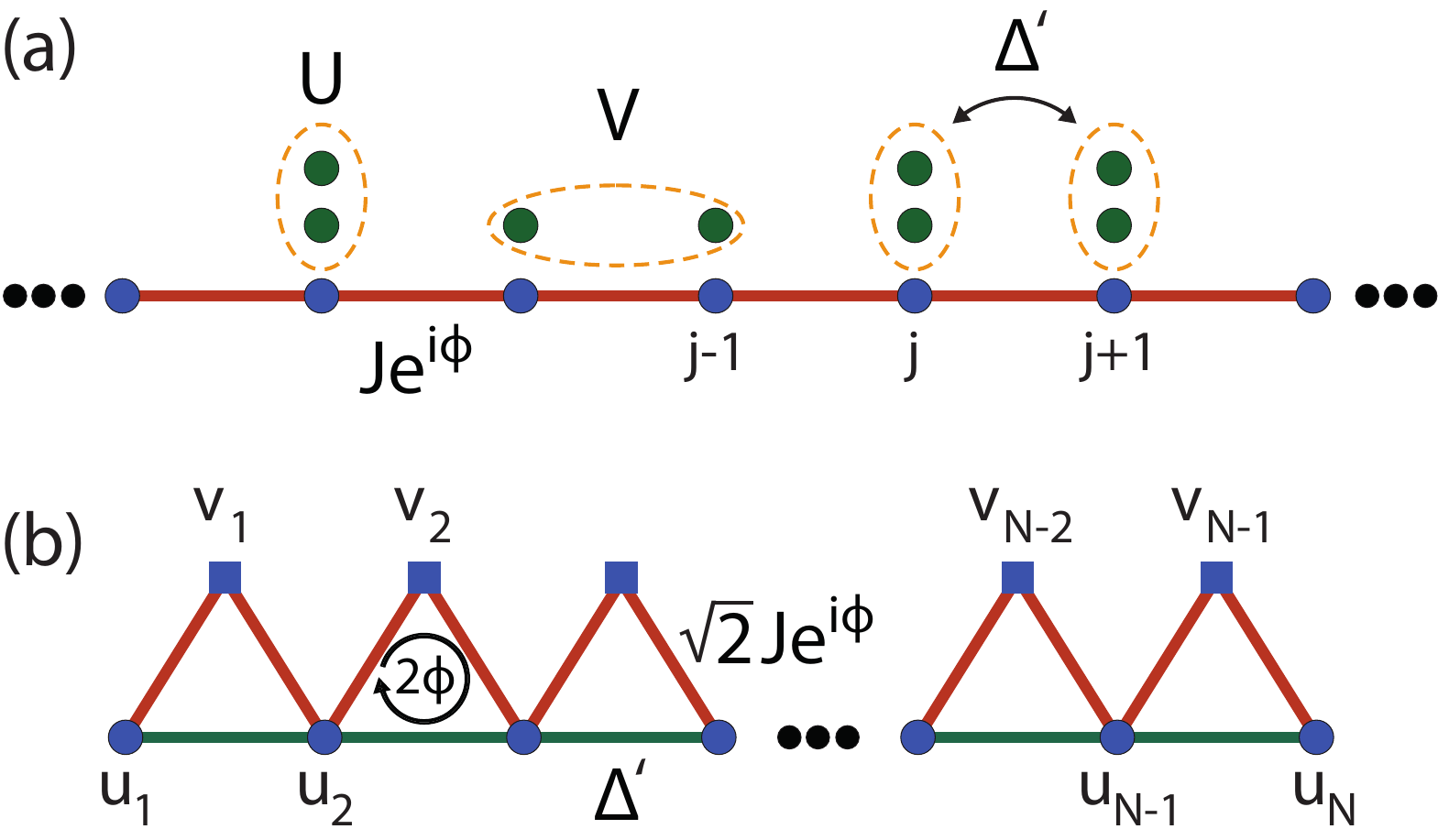} 
		\par\end{centering}
	\caption{(a) Linear open chain with Hubbard ($U$) and NN ($V$) interactions, supplemented with pair hopping terms ($\Delta^\prime$). A Peierls phase $\phi$ is ascribed to each coupling term $J$. (b) Effective model for the HES, consisting of a sawtooth chain with a $2\phi$ flux per plaquette and onsite energy $U$ ($V$) at the $u_j$ ($v_j$) sites.}
	\label{fig:extbosehub}
\end{figure}

Let us consider the Hamiltonian of the EBH model depicted in Fig.~\ref{fig:extbosehub}(a),
\begin{eqnarray}
	H_{\text{EBH}}&=&H_{\text{int}}+H_{\text{kin}},
	\label{eq:hamiltebh}
	\\
	H_{\text{int}}&=&\frac{U}{2}\sum\limits_{j=1}^N n_j(n_j-1) + V\sum\limits_{j=1}^{N-1}n_jn_{j+1},
	\\
	H_{\text{kin}}&=&\sum\limits_{j=1}^{N-1}\Big[\frac{\Delta^\prime}{2} b_{j+1}^\dagger b_{j+1}^\dagger b_j b_j -Je^{i\phi}b_{j+1}^\dagger b_j + H.c.\Big],
\end{eqnarray}
where $N$ is the number of sites of the open chain, $H_{\text{int}}$ is the interacting part, with $U$ the strength of the Hubbard interaction and $V$ of the NN interaction, $n_j=b_j^\dagger b_j$ is the number operator with $b_j$ the bosonic annihilation operator acting on site $j$, $H_{\text{kin}}$ is the kinetic part, composed both of single-particle hoppings with magnitude $J$, set henceforth as the energy unit, and phase $\phi$ and of pair hoppings between NN sites of strength $\Delta^\prime$.
While the strength of the interactions is assumed constant in this work, we note that recent studies have addressed the emergent topological effects that result from modulating the Hubbard term in two-body systems, both in the weak \cite{Pelegri2020,Kuno2020a} and strong \cite{Huang2024} interactions regime.

In the strong interactions limit ($U,V\gg J,\Delta^\prime$), the Hilbert space $\mathcal{H}
$ can be divided in two:

(i) High-energy subspace (HES). This Hilbert subspace reads as $\mathcal{H}_{\text{HES}}=\{\{\ket{u_i}\},\{\ket{v_j}\}\}$, with $\ket{u_i}=\frac{b_i^\dagger b_i^\dagger}{\sqrt{2}}\ket{\emptyset}$, $i=1,2,\dots,N$ and $\ket{\emptyset}$ the vacuum state, corresponding to doublon states \cite{Winkler2006,Creffield2010,Salerno2020,Iskin2023,Nicolau2023,Alyuruk2024,Pelegri2024} where the two bosons occupy the same site, and $\ket{v_j}=b_j^\dagger b_{j+1}^\dagger\ket{\emptyset}$, with $j=1,2,\dots,N-1$, corresponding to states with the two-particles on NN sites.
Their interaction energies are given by
\begin{eqnarray}
	\bra{u_i}H_{\text{int}}\ket{u_i}&=&U,\ \  \forall i,
	\\
	\bra{v_j}H_{\text{int}}\ket{v_j}&=&V,\ \  \forall j.
\end{eqnarray}
The dimension of this subspace is $\dim \mathcal{H}_{\text{HES}}=2N-1$.

(ii) Low-energy subspace (LES). This Hilbert subspace reads as $\mathcal{H}_{\text{LES}}=\{\ket{i,j}\}$, where $\ket{i,j}=b_i^\dagger b_j^\dagger\ket{\emptyset}$, with $i=1,2,\dots,N-2$ and $j=i+2,i+3,\dots,N$, that is, the two particles in the states of this subspace are separated by at least two sites ($j-i\geq 2$).
The interaction energy of these states is given by
\begin{eqnarray}
	\bra{i,j}H_{\text{int}}\ket{i,j}=0,\ \  \forall i,j.
\end{eqnarray}
The dimension of this subspace is $\dim \mathcal{H}_{\text{LES}}={N \choose 2} - N+1$, where the subtracted $N-1$ states correspond to the number of states in the set $\{\ket{v_j}\}$.

With $H_{\text{kin}}$ assumed to be a perturbation, the non-zero first-order corrections in HES are given by
\begin{eqnarray}
		\bra{v_j}H_{\text{kin}}\ket{u_j}&=&\bra{u_{j+1}}H_{\text{kin}}\ket{v_j}=-\sqrt{2}Je^{i\phi},
		\\
		\bra{u_j}H_{\text{kin}}\ket{u_{j+1}}&=&\Delta^\prime,
\end{eqnarray}
which can be regarded as the couplings of the effective single-particle HES model, with a Hamiltonian given by
\begin{widetext}
\begin{equation}
	H_{\text{HES}}=\Delta^\prime\sum\limits_{j=1}^{N-1}\big(u_j^\dagger u_{j+1}+H.c\big) - \sqrt{2}J\sum_{j=1}^{N-1}\Big(v_{j}^\dagger(e^{i\phi}u_j+e^{-i\phi}u_{j+1}) +H.c\Big) + U\sum\limits_{j=1}^N u_j^\dagger u_j + V\sum\limits_{j=1}^{N-1}v_j^\dagger v_j,
\end{equation}
\end{widetext}
where we have redefined $\ket{u_j}\equiv u_j^\dagger\ket{\emptyset}$ and $\ket{u_j}\equiv v_j^\dagger\ket{\emptyset}$ as effective single-particle states.
This Hamiltonian models the sawtooth chain depicted in Fig.~\ref{fig:extbosehub}(b).
Crucially, the Peierls phases of the original open chain carry over to the effective model, generating a $2\phi$ flux per plaquette.
% Figure
%%%%%%%%%%%%%%%%%%%%%%%%%%%%%%%%%%%%%%%%%%%%%%%%%%%%%%%%%%%%%%%%%%%%%
%%%%%%%%%%%%%%%%%%%%%%%%%%%%%%%%%%%%%%%%%%%%%%%%%%%%%%%%%%%%%%%%%%%%%
%%%%%%%%%%%%%%%%%%%%%%%%%%%%%%%%%%%%%%%%%%%%%%%%%%%%%%%%%%%%%%%%%%%%%
\begin{figure*}[ht]
	\begin{centering}
		\includegraphics[width=0.98 \textwidth,height=6.6cm]{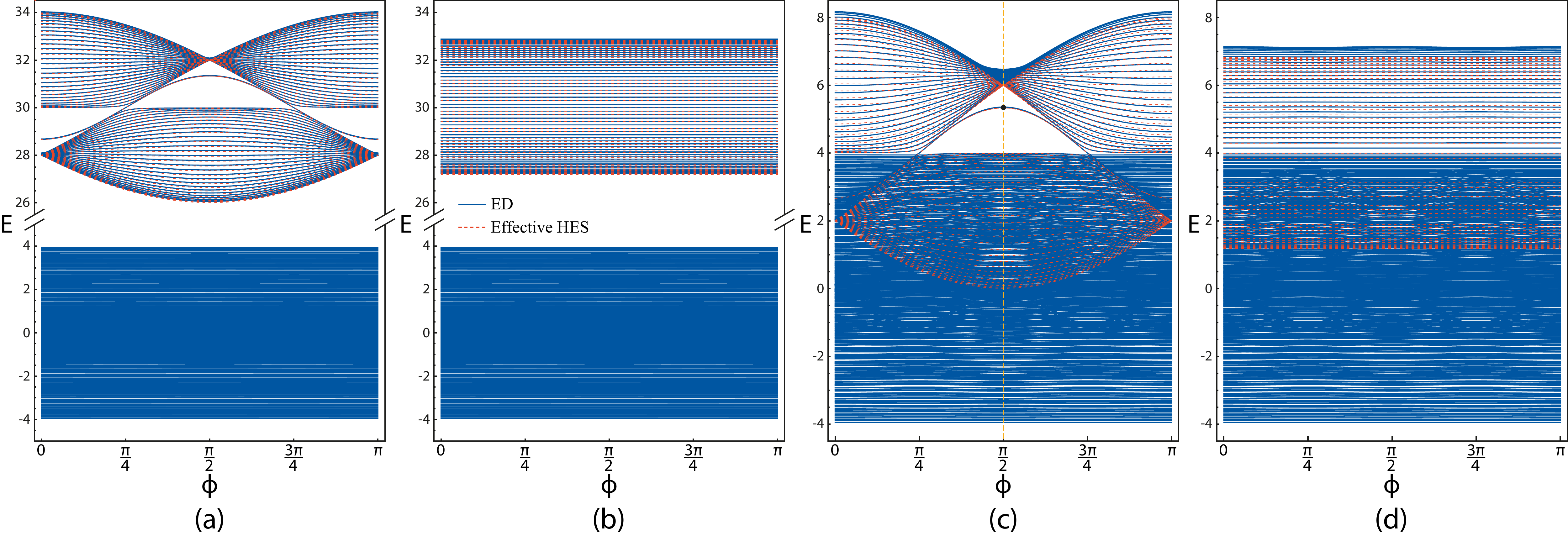} 
		\par\end{centering}
	\caption{Energy, in units of $J$, as a function of the Peierls phase $\phi$ of the open EBH chain with $N=30$ sites for (a) $U=V=30$ and $\Delta^\prime=1$, (b) $U=V=30$ and $\Delta^\prime=0$, (c) $U=V=4$ and $\Delta^\prime=1$, and (d) $U=V=4$ and $\Delta^\prime=0$. Solid blue curves come from exact diagonalization (ED) of the model in Fig.~\ref{fig:extbosehub}(a), while dashed red curves come from diagonalization of the effective HES model in Fig.~\ref{fig:extbosehub}(b). Note the break in the energy axis in (a) and (b).}
	\label{fig:espectrumebh}
\end{figure*}
%%%%%%%%%%%%%%%%%%%%%%%%%%%%%%%%%%%%%%%%%%%%%%%%%%%%%%%%%%%%%%%%%%%%%
%%%%%%%%%%%%%%%%%%%%%%%%%%%%%%%%%%%%%%%%%%%%%%%%%%%%%%%%%%%%%%%%%%%%%
%%%%%%%%%%%%%%%%%%%%%%%%%%%%%%%%%%%%%%%%%%%%%%%%%%%%%%%%%%%%%%%%%%%%%
For $U=V$ and $\Delta^\prime=J$, the sawtooth chain is known \cite{Zhang2015,Weimann2016,Kuno2020,Yang2021} to exhibit a lowest (highest) energy flat band for $2\phi=0$ ($2\phi=\pi$) gapped from a second dispersive band, along with in-gap edge states for open boundaries.  

In Fig.~\ref{fig:espectrumebh}, the energy spectrum from exact diagonalization (ED) of the EBH open chain of Fig.~\ref{fig:extbosehub}(a) for $N=30$ sites, together with that of the respective effective HES model in Fig.~\ref{fig:extbosehub}(b), is plotted as a function of the Peierls phase for four different sets of parameters. 
For the strong interactions regime, in Figs.~\ref{fig:espectrumebh}(a) and \ref{fig:espectrumebh}(b), a very good agreement is found at high energies between the ED and the effective HES results.
The LES spectrum of both these figures can be approximately viewed as that of the energy sum of two independent particles states on a linear open chain, such that the Peierls phase $\phi$ at the hopping parameters is inconsequential.
The HES spectrum in Fig.~\ref{fig:espectrumebh}(a) has gapped regions traversed by a doubly degenerate (up to negligible finite-size corrections) curve of in-gap edge states, as well as a clear dependence on the Peierls phase,  which in particular can be responsible for gap closings and reopenings when varied adiabatically.
Conversely, no $\phi$ dependence is observed in the HES of Fig.~\ref{fig:espectrumebh}(b).
This can be understood in terms of the effective HES model in Fig.~\ref{fig:extbosehub}(b) since, when $\Delta^\prime=0$, the sawtooth chain with closed triangle loops pierced by a $2\phi$  magnetic flux reduces to a linear open chain, and therefore the Peierls phases can be gauged away. 

In Fig.~\ref{fig:espectrumebh}(c), we are no longer in the strong interactions regime since $U=V=4$.
However, good agreement between the ED and the effective HES results can still be found for the higher energy states above the lower continuum ($E>4$), such that the latter still capture the essential features of the model in this region.
In particular, while there is no energy gap at $\phi=0$, a gap hosting a doubly degenerate band of edge states is present above $E=4$ within the $\frac{\pi}{4}\lesssim\phi\lesssim\frac{3\pi}{4}$ region, with a maximum at $\frac{\pi}{2}$ [along the vertical dashed orange line in Fig.~\ref{fig:espectrumebh}(c)].
This choice of parameters highlights the non-trivial effects of including the Peierls phase at the hopping parameters of the linear EBH open chain, since the edge states only become visible in a finite region of $\phi$ values and are buried in the LES continuum of states for $\phi=0$.
We will take advantage of this in the next section, when we map exactly the EBH chain onto an electrical circuit in the equivalent $\phi=\frac{\pi}{2}$ regime.
In Fig.~\ref{fig:espectrumebh}(d), we used the same parameters as in Fig.~\ref{fig:espectrumebh}(c), except for setting now $\Delta^\prime=0$ such that, as with the case of Fig.~\ref{fig:espectrumebh}(b), no $\phi$ dependence is observed on the spectrum for $E>4$ (up to possible negligible higher-order corrections), meaning that the same qualitative picture holds here, namely that a trivial $\Delta^\prime$ in the effective chain of Fig.~\ref{fig:extbosehub}(b) neutralizes the effect of $\phi$.

A general two-boson state of the model with the Hamiltonian $H_{\text{EBH}}$ in (\ref{eq:hamiltebh}) can be written as
\begin{equation}
	\ket{\psi}=\sum\limits_{j,l=1}^N\frac{\psi_{j,l}}{\tilde{\delta}_{j,l}}b^\dagger_jb^\dagger_l\ket{\emptyset},
	\label{eq:psi}
\end{equation}
where $\tilde{\delta}_{j,l}:=1+(\sqrt{2}-1)\delta_{j,l}$, together with the constraint $\psi_{j,l}=\psi_{j,l}$.
From the Schr\"odinger equation $H_{\text{EBH}}\ket{\psi}=E\ket{\psi}$ we can extract the tight-binding equation for every bulk component, that is, for $j,l=2,3,\dots,N-1$ (the general equation, including the boundary components, is given in Appendix~\ref{app:boundaryeigen}),
\begin{widetext}
	\begin{eqnarray}
		E\psi_{j,l}=&-&J\big[e^{i\phi}(\psi_{j-1,l}+\psi_{j,l-1}) + e^{-i\phi}(\psi_{j+1,l}+\psi_{j,l+1})\big]   \nonumber
		\\
		&+& \Delta^\prime \delta_{j,l} \big[\psi_{j-1,l-1}+\psi_{j+1,l+1}\big] + U\delta_{j,l}\psi_{j,l} + V\delta_{j,l\pm 1}\psi_{j,l}.
		\label{eq:eigeneqebh}
	\end{eqnarray}
\end{widetext}
While this is the eigenequation for two particles in a 1D chain, with one defined along the $j$-coordinate and the other along the $l$-coordinate, several studies have shown that it can be exactly mapped onto a single-particle 2D system, with the particle living on the lattice generated by the all the possible $(j,l)$ coordinates \cite{Valiente2008,Longhi2011,Krimer2011,Corrielli2013,Mukherjee2016,DiLiberto2016,Gorlach2017a,Gorlach2018}.
Under this mapping, U and V interactions translate as diagonal and off-diagonal onsite potentials \cite{Gorlach2017}, while $\Delta^\prime$ translates as a diagonal coupling connecting adjacent diagonal sites.

If we properly symmetrize the components in (\ref{eq:psi}) by defining
\begin{equation}
	\begin{cases}
		\psi_{j,l}^B:=\frac{1}{\sqrt{2}}(\psi_{j,l}+\psi_{l,j}),\ \  \text{for}\  j<l,
		\\
		\psi_{j,l}^B:=\psi_{j,l},\ \  \text{for}\  j=l,
	\end{cases}
\label{eq:bosonization}
\end{equation}
then (\ref{eq:eigeneqebh}) can be recast as
\begin{widetext}
\begin{eqnarray}
	E\psi_{j,l}^B=&-&J\tilde{\delta}_{j,l}\big[e^{i\phi}\psi_{j-1,l}^B+e^{-i\phi}\psi_{j,l+1}^B+\tilde{\delta}_{j,l-1}\bar{\delta}_{j,l}(e^{i\phi}\psi_{j,l-1}^B+e^{-i\phi}\psi_{j+1,l}^B)\big]   \nonumber
	\\
	&+& \Delta^\prime\delta_{j,l}\big[\psi_{j-1,l-1}^B+\psi_{j+1,l+1}^B\big]+U\delta_{j,l}\psi_{j,l}^B + V\delta_{j,l-1}\psi_{j,l}^B,
	\label{eq:eigeneqebhbos}
\end{eqnarray}
\end{widetext}
where we introduced the complementary Kronecker delta $\bar{\delta}_{n,m}:=1-\delta_{n,m}$, while the $\tilde{\delta}_{n,m}$ serve to introduce a $\sqrt{2}$ factor in the couplings connected to the doubly occupied states.

% SECTION
%%%%%%%%%%%%%%%%%%%%%%%%%%%%%%%%%%%%%%%%%%%%%%%%%%%%%%%%%%%%%%%%%%%%%
%%%%%%%%%%%%%%%%%%%%%%%%%%%%%%%%%%%%%%%%%%%%%%%%%%%%%%%%%%%%%%%%%%%%%
%%%%%%%%%%%%%%%%%%%%%%%%%%%%%%%%%%%%%%%%%%%%%%%%%%%%%%%%%%%%%%%%%%%%%
\section{Electrical circuit implementation of the EBH chain}
\label{sec:lccircuit}
For a given LC circuit, the response of the system to an applied voltage of angular frequency $\omega$ \cite{Lee2018,Dong2021,Yang2024} can be codified as 
\begin{equation}
	I_a(\omega)=\sum\limits_bA_{ab}(\omega)\mathcal{V}_b(\omega),
	\label{eq:circuiteq}
\end{equation}
where $I_a$ and $\mathcal{V}_a$ denote the input current and voltage at node $a$, which are related by the grounded circuit Laplacian $A$, whose components have the general form
\begin{eqnarray}
	A_{aa}(\omega)&=&i\omega\Big[ C_a-\frac{1}{\omega^2L_{a}}+\sum\limits_{b\neq a}\Big( C_{ab}-\frac{1}{\omega^2L_{ab}}\Big)\Big],
		\\
		A_{ab}(\omega)&=&i\omega\Big[\frac{1}{\omega^2L_{ab}}-C_{ab}\Big],\ \ a\neq b,
\end{eqnarray}
where $L_a$ and $L_{ab}=L_{ba}$ ($C_a$ and $C_{ab}=C_{ba}$) is the inductance (total capacitance) of the inductor (capacitors) coupling node $a$ to the ground and to node $b$, respectively.
As such, we are assuming that only one inductor $L_a$ ($L_{ab}$) can couple node $a$ to the ground (node $b$), while $C_a$ ($C_{ab}$) is the total capacitance of an arbitrary number of capacitors coupled in parallel to the ground (node $b$).
By writing (\ref{eq:circuiteq}) in matrix notation, 
\begin{equation}
	\mathbf{I}=A\mathbf{\mathcal{V}}=j(\omega)\mathbf{\mathcal{V}},
	\label{eq:circuiteqmatrix}
\end{equation}
where $j(\omega)$ is the admittance, it becomes clear that a formal equivalency can be made between (\ref{eq:circuiteqmatrix}) and the Schr\"odinger equation $H\ket{\psi}=E\ket{\psi}$ of a single-particle system, with the set of correspondences $\{A,\mathbf{\mathcal{V}},j(\omega)\}\to\{H,\ket{\psi},E\}$.
% Figure
%%%%%%%%%%%%%%%%%%%%%%%%%%%%%%%%%%%%%%%%%%%%%%%%%%%%%%%%%%%%%%%%%%%%%
%%%%%%%%%%%%%%%%%%%%%%%%%%%%%%%%%%%%%%%%%%%%%%%%%%%%%%%%%%%%%%%%%%%%%
%%%%%%%%%%%%%%%%%%%%%%%%%%%%%%%%%%%%%%%%%%%%%%%%%%%%%%%%%%%%%%%%%%%%%
\begin{figure*}[ht]
	\begin{centering}
		\includegraphics[width=0.95 \textwidth,height=6.95cm]{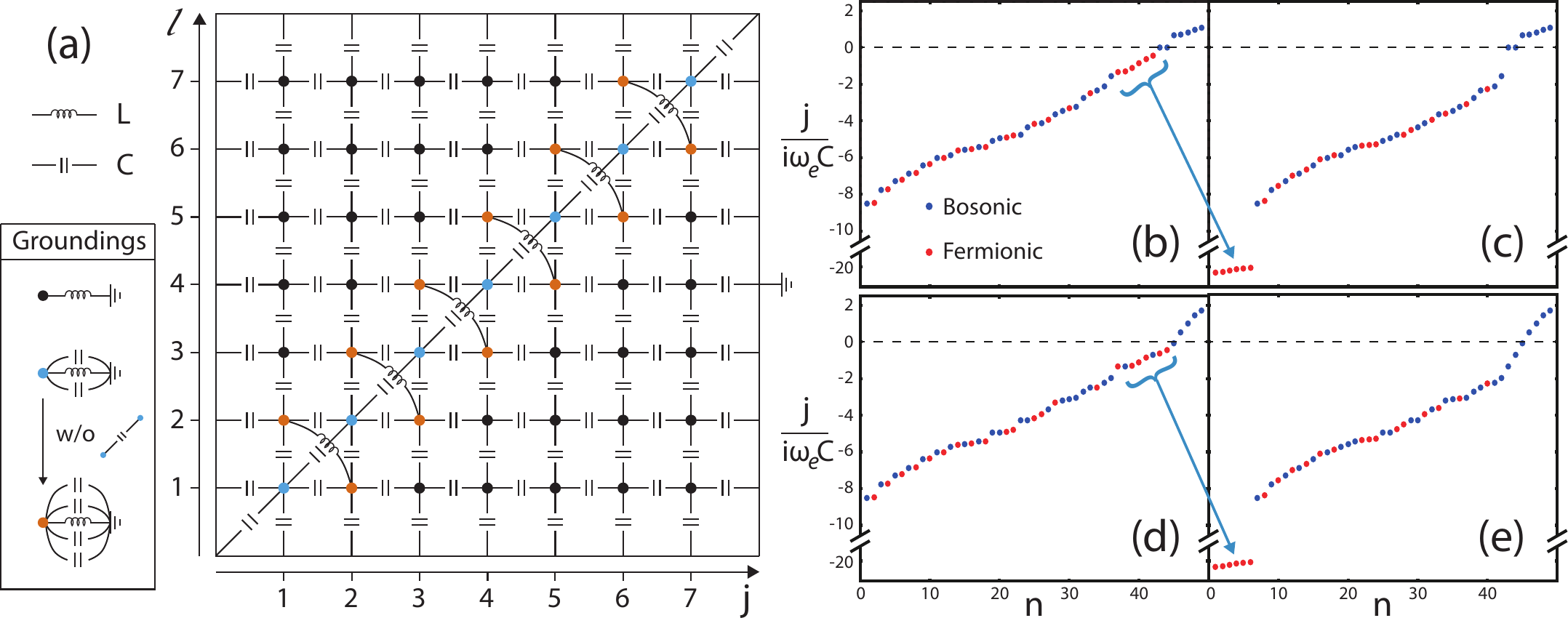} 
		\par\end{centering}
	\caption{(a) Electrical circuit lattice simulation of the two-body EBH model in Fig.~\ref{fig:extbosehub}(a) with $N=7$ sites, $U=V=4$ and $\Delta^\prime=1$ ($\Delta^\prime=0$) when the diagonal capacitors coupling blue nodes are present (absent).
		Admittance spectrum of the circuit in (a) at the resonant angular frequency $\omega_e$ with diagonal capacitors and (b) without diagonal inductors, and (c) with diagonal inductors coupling orange nodes.
		(d)-(e) Same as in (b)-(c), respectively, but without diagonal capacitors (and with the corresponding change in the grounding scheme of the blue nodes).
		Blue (red) states in (b)-(e) are labeled as bosonic (fermionic) states due to their symmetric (anti-symmetric) voltage profile in relation to the main diagonal in (a).
		Note the break in the admittance axis in (b)-(e).}
	\label{fig:eleccirc}
\end{figure*}
%%%%%%%%%%%%%%%%%%%%%%%%%%%%%%%%%%%%%%%%%%%%%%%%%%%%%%%%%%%%%%%%%%%%%
%%%%%%%%%%%%%%%%%%%%%%%%%%%%%%%%%%%%%%%%%%%%%%%%%%%%%%%%%%%%%%%%%%%%%
%%%%%%%%%%%%%%%%%%%%%%%%%%%%%%%%%%%%%%%%%%%%%%%%%%%%%%%%%%%%%%%%%%%%%

As mentioned at the end of the previous section, the two-bosons EBH $N$-sites chain depicted in Fig.~\ref{fig:extbosehub}(a) can be mapped onto a single-particle 2D system of $N\times N$ sites which, in turn, has an electric circuit analog as we have shown above.
For the grounded circuit depicted in Fig.~\ref{fig:eleccirc}(a) with $N\times N$ nodes, with $N=7$, we show in Appendix~\ref{app:circuiteigen} that the bulk eigenequation, defined for $j,l=2,3,\dots,N-1$, for the symmetric modes with respect to the main diagonal, which we label bosonic modes with components $\mathcal{V}^B_{j,l}$, can be written as
\begin{widetext}
	\begin{eqnarray}
		j^\prime(\omega)\mathcal{V}^B_{j,l}=&-&i\tilde{\delta}_{j,l}\big[\mathcal{V}^B_{j-1,l}-\mathcal{V}^B_{j,l+1}+\tilde{\delta}_{j,l-1}\bar{\delta}_{j,l}(\mathcal{V}^B_{j,l-1}-\mathcal{V}^B_{j+1,l})\big] \nonumber
		\\
		&+&\delta_{j,l}(\mathcal{V}^B_{j-1,l-1}+\mathcal{V}^B_{j+1,l+1}) + 4\delta_{j,l}\mathcal{V}^B_{j,l}+4\delta_{j,l-1}\mathcal{V}^B_{j,l}, 
		\label{eq:eigencircuitbosmain}
	\end{eqnarray}
\end{widetext}
with $j^\prime(\omega)=\big(\frac{\omega_0^2}{\omega^2}-4\big)=\big(\frac{f_0^2}{f^2}-4\big)$, $\omega=2\pi f$ and $\omega_0=\sqrt{\frac{1}{LC}}$ the natural angular frequency of the circuit.
From comparing (\ref{eq:eigeneqebhbos}) with (\ref{eq:eigencircuitbosmain}), it is clear that the former can be exactly mapped onto the latter according to the following set of correspondences and parameters,
\begin{equation}
	\begin{cases}
		\psi^B_{j,l}\to\mathcal{V}_{j,l}^B,
		\\
		E\to j^\prime(\omega),
		\\
		J=\Delta^\prime =1,
		\\
		\phi=\frac{\pi}{2},
		\\
		U=V=4.
		\label{eq:ebhtocircuit}
	\end{cases}
\end{equation}
For the EBH chain with $N=30$ sites, these parameters yield the energy spectrum of Fig.~\ref{fig:espectrumebh}(c) along the vertical orange dashed line where, in particular, a pair of degenerate (up to negligible corrections due to finite-size effects) edge states, marked by the black dot, appear at the energy gap.
Below, we will be interested in searching for signatures of these states in their electrical circuit equivalents, since their presence can be attributed, in the original EBH model, to the effect of the Peierls phase $\phi$, given the energy gap is absent for $\phi=0$ and/or for $\Delta^\prime=0$, as shown in Fig.~\ref{fig:espectrumebh}(d).

However, it is important to notice that the exact mapping defined through (\ref{eq:ebhtocircuit}) is an injective mapping, meaning that there is a one-to-one correspondence between each eigenstate of the EBH chain and each voltage eigenstate of the electrical circuit, but only within the symmetric bosonic sector of the circuit.
The complete admittance spectrum of the circuit also has extra states whose voltage profile is antisymmetric with respect to the main diagonal, which we label (spinless) fermionic modes with components $\mathcal{V}^F_{j,l}$, since double occupancy is forbidden, that is, there are zeros of voltage at the nodes along the main diagonal.
As shown in Appendix~\ref{app:circuiteigen}, the eigenequation for the bulk fermionic modes of the circuit in Fig.~\ref{fig:eleccirc}(a), defined for $l=3,4,\dots,N-1$ and $j=2,3,\dots,l-1$, is given by
\begin{widetext}
	\begin{equation}
		j^\prime(\omega)\mathcal{V}^F_{j,l}=-\big[\mathcal{V}^F_{j-1,l}+\mathcal{V}^F_{j,l+1}+\bar{\delta}_{j,l-1}(\mathcal{V}^F_{j,l-1}+\mathcal{V}^F_{j+1,l})\big]+V(f)\delta_{j,l-1}\mathcal{V}^F_{j,l},
		\label{eq:eigencircuitfermain}
	\end{equation}
\end{widetext}
where we now have a frequency-dependent onsite potential $V(f)=4-2\frac{f_0^2}{f^2}$ (the circuit equivalent of NN interactions) at the off-diagonal nodes with $j=l-1$, whose last term $-2\frac{f_0^2}{f^2}$ originates from the inductors coupling the orange nodes on both sides of the main diagonal in Fig.~\ref{fig:eleccirc}(a).
Since this term is absent from (\ref{eq:eigencircuitbosmain}), it allows us to control the spectrum of the fermionic sector without affecting the bosonic one.
The effect of removing the diagonal inductors translates as setting $V(f)=4$ in (\ref{eq:eigencircuitfermain}), with everything else remaining the same.

In Fig.~\ref{fig:eleccirc}(b), we plot the admittance spectrum of the circuit in Fig.~\ref{fig:eleccirc}(a) without the diagonal inductors, distinguishing between the bosonic modes (blue dots) and fermionic modes (red dots).
The driving frequency corresponds to the angular eigenfrequency of the degenerate (up to negligible finite-size corrections) pair of gapped bosonic modes with value $w_e\approx 0.327\omega_0$, yielding $j^\prime(\omega_e)\approx 5.358$, which coincides with the energy value of the edge states in Fig.~\ref{fig:espectrumebh}(c), as expected, while $j(\omega_e)=0$ in (\ref{eq:eigencircuit}), that is, $\omega_e$ is the resonant angular frequency at which the admittance of the in-gap states rests at the zero level.
The bosonic admittance gap below the zero-energy modes is filled with close-by fermionic modes, all of whom with large weight on the off-diagonal orange nodes in Fig.~\ref{fig:eleccirc}(a), i.e., these are the fermionic states most affected by $V$.
When we include the diagonal inductors, only the fermionic sector of the admittance spectrum changes, as shown in Fig.~\ref{fig:eleccirc}(c) (conversely, if the diagonal couplings were composed of capacitors instead of inductors, only the symmetric bosonic sector would be affected, as shown in \cite{Wang2023}).
In particular, since we now have $V(f_e)=-14.716$, where $f_e\approx 0.327f_0$, the high-admittance fermionic modes in Fig.~\ref{fig:eleccirc}(b) become the lowest admittance bundle in Fig.~\ref{fig:eleccirc}(c), that is, they leave the bosonic gap, such that no spurious modes surround the targeted zero-admittance modes.

In Fig.~\ref{fig:eleccirc}(d), we plot the admittance spectrum of the circuit in Fig.~\ref{fig:eleccirc}(a) without both the diagonal inductors and the diagonal capacitors, while grounding the blue nodes the same way as the orange nodes, so as to keep $U=4$ unchanged, which corresponds, in the EBH chain, to the $\Delta^\prime=0$ case [see the energy spectrum in Fig.~\ref{fig:espectrumebh}(d) for $\phi=\frac{\pi}{2}$].
The driving angular frequency is kept at $\omega_e$.
As expected, there is no bosonic admittance gap in this case around the zero level, however some fermionic modes are still present in this region.
As before, we send them down in the spectrum by including the diagonal inductors in the circuit, as shown in Fig.~\ref{fig:eleccirc}(e).

%%%%%%%%%%%%%%%%%%%%%%%%%%%%%%%%%%%%%%%%%%%%%%%%%%%%%%%%%%%%%%%%%
%%%%%%%%%%%%%%%%% Figure %%%%%%%%%%%%%%%%%%%%%%%%%%%%%%%%%%%%%%%%
%%%%%%%%%%%%%%%%%%%%%%%%%%%%%%%%%%%%%%%%%%%%%%%%%%%%%%%%%%%%%%%%%
\begin{figure}[ht]
	\begin{centering}
		\includegraphics[width=0.45\textwidth]{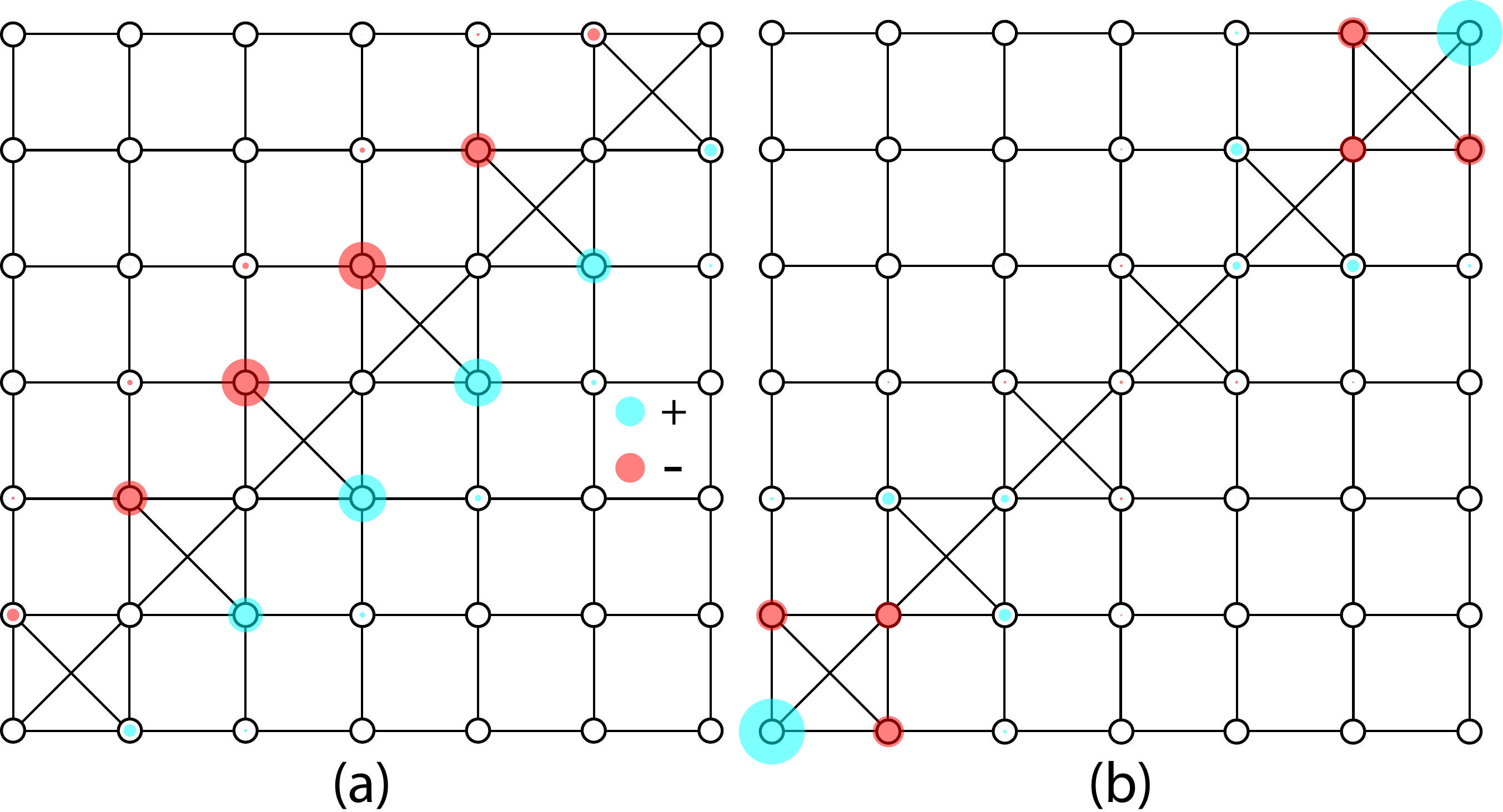} 
		\par\end{centering}
	\caption{Voltage profile at the nodes (white circles) of the 2D electrical circuit in Fig.~\ref{fig:eleccirc}(a) for the (a) lowest eigenstate and (b) right eigenstate of the pair resting at zero admittance in Fig.~\ref{fig:eleccirc}(c), for the same parameters as in there. The radius of the circle represents the voltage amplitude at the node, while the color indicates the respective sign.}
	\label{fig:voltage}
\end{figure}
The voltage profile of the lowest eigenstate in Fig.~\ref{fig:eleccirc}(c) is shown in Fig.~\ref{fig:voltage}(a).
This is a 1D state exponentially decaying from the off-diagonal nodes connected by the extra inductors, where most of its weight is concentrated, and anti-symmetric about the diagonal line of symmetric nodes, where it has no weight, showing that this state corresponds to a fermionic mode.
In Fig.~\ref{fig:voltage}(b), on the other hand, we represent the voltage function of the right eigenstate of the pair at the zero-admittance level in Fig.~\ref{fig:eleccirc}(c).
It can be seen that this is a zero-dimensional (0D) state decaying from opposite corners symmetrically about the diagonal line of symmetric nodes, which corresponds to a bosonic mode.
Since the electrical circuit of Fig.~\ref{fig:eleccirc}(a) represents a mapping of the open 1D interacting system in Fig.~\ref{fig:extbosehub}(a), this corner voltage state is the electrical counterpart of the in-gap state at $\phi=\frac{\pi}{2}$ in Fig.~\ref{fig:espectrumebh}(c).
Given that the mapping is \textit{exact}, the experimental detection of the corner state would serve as demonstration of the physical effects of the Peierls phase $\phi$ in an interacting model built on an open linear chain.

Experimental realization of the electrical circuit in Fig.~\ref{fig:eleccirc}(a) should be straightforward, since only one type of capacitor, with capacitance $C$, and one type of inductor, with inductance $L$, were considered here.
Low-tolerance components should be considered to avoid closing the gap around zero admittance in the spectrum [see Fig.~\ref{fig:eleccirc}(c)].
%This also explains why we have considered the $U=V=4$ case which, strictly speaking, is far from the strong interactions limit, since $U,V\gg 1$ would require, in the mapped electrical circuit, a large grounding capacitance for the blue and orange nodes in Fig.~\ref{fig:eleccirc}(c).
%In this case, the tolerance-induced disorder emanating from the grounding components at these nodes is not expected to be negligible, when compared with the relevant admittance gap at the HES. 
While we focused on the $U=V=4$ case, we note, however, that in a study on a similar lattice circuit \cite{Zhang2021}, and using circuit elements with $1\%$ tolerance, the authors were able to experimentally probe the HES with $U$ and $V$ up to $15$ each.
The system is then fed with an alternating current of frequency $f_e$, determined by the $C$ and $L$ values, bringing the corner modes in the bosonic gap in Fig.~\ref{fig:eleccirc}(c) into resonance, creating a peak in the two-point impedance \cite{Lee2018}, provided the voltage drop between the two nodes considered for the measurement is maximized [e.g., for the corner mode in Fig.~\ref{fig:voltage}(b), one of the nodes should be a corner one where there is a maximum amplitude, and the other a bulk node with zero amplitude].
Finally, with voltage measurements at each node one can probe the entire lattice to find the voltage profile of the corner modes \cite{Olekhno2020}, whose hybridization due to finite size effects can be prevented by perturbing one of the main diagonal corners of the lattice by, e.g., grounding a corner node with extra capacitors or suppressing it altogether.

% SECTION
%%%%%%%%%%%%%%%%%%%%%%%%%%%%%%%%%%%%%%%%%%%%%%%%%%%%%%%%%%%%%%%%%%%%%
%%%%%%%%%%%%%%%%%%%%%%%%%%%%%%%%%%%%%%%%%%%%%%%%%%%%%%%%%%%%%%%%%%%%%
%%%%%%%%%%%%%%%%%%%%%%%%%%%%%%%%%%%%%%%%%%%%%%%%%%%%%%%%%%%%%%%%%%%%%
\section{Conclusions}
\label{sec:conclusions}
For a certain class of tight-binding systems, the inclusion of Peierls phases in the hopping terms can have non-trivial effects even when the underlying model in real-space is given by a linear open chain, such as the Kitaev and EBH models studied here.
This somewhat counter-intuitive picture is a consequence of the fact that closed paths can emerge in the adjacency graph of the Fock space, around which Peierls phases can accumulate to generate a magnetic flux analog, even though closed paths might be absent in real-space. 
%(or, more generally, at the single-particle level).
An inverse approach to the one followed here can be applied to future studies in many-body systems: upon construction of the many-body Fock space graph \cite{Santos2019,Balas2020,Santos2019,Roy2019,Roy2020}, one can ask how the introduction of a flux crossing a given closed path in this space would translate in terms of the real-space model, that is, what kind of phase factors would have to be included in the original Hamiltonian in order to produce such fluxes in the Fock space.

The independent control over the Peierls phases picked up along the single-particle hopping terms can become a useful tool when applied to the Kitaev model. 
Specifically, it was shown that a patterned arrangement of these phases can lead to the appearance of Majorana flat bands and also topological states in the Majorana spectrum.
Local superconducting double $\pi$-junctions can also be emulated by tuning the signs of the hopping parameters at specific locations of the chain.
An interesting open question relates to the possibly of directing MZMs across the chain via successive flippings of the hopping signs.

Going beyond quadratic models, we proceeded to study an open EBH linear chain \cite{Zeybek2024}, including a pair hopping term and finite Peierls phases at the single-particle hopping term.
By restricting our analysis to a two-body system, it was shown that the HES of the strong interactions regime can be understood, to first order, as the spectrum of single-particle states in an effective sawtooth chain, depicted in Fig.~\ref{fig:extbosehub}(b), where the original Peierls phases generate an effective magnetic flux at each plaquette.
For an appropriately chosen parameter set, there is flux interval for which in-gap edge states appear between the HES and LES continua of states [see Fig.~\ref{fig:espectrumebh}(c)], highlighting the physical effects of the Peierls phases in this open linear model, which are fundamentally distinct from those produced when these phases either reflect synthetic or real fluxes \cite{Zheng2023}, or derive from generalized anyon statistics \cite{Olekhno2022}.  

Finally, after showing that the two-body EBH chain can be recast as a single-particle system on a square lattice, we provided an exact mapping of the latter onto an equivalent electrical circuit \cite{Olekhno2020,Zhang2021}.
The circuit components can be selected in a manner that emulates the EBH chain with a parameter set, which includes finite Peierls phases, for which in-gap states are predicted to be present.
Numerical results on the lattice circuit confirmed the presence of the mapped in-gap corner states in the admittance spectrum [see Fig.~\ref{fig:eleccirc}(c) and Fig.~\ref{fig:voltage}(b)].
As such, the implementation of the lattice circuit introduced in Sec.~\ref{sec:lccircuit}, together with the detection of these corner voltage modes, is a tabletop solution that offers a simple and easy path for the experimental observation of the effects that Peierls phases can induce even in systems built on open linear chains.

% Appendices
%%%%%%%%%%%%%%%%%%%%%%%%%%%%%%%%%%%%%%%%%%%%%%%%%%%%%%%%%%%%%%%%%%%%%
%%%%%%%%%%%%%%%%%%%%%%%%%%%%%%%%%%%%%%%%%%%%%%%%%%%%%%%%%%%%%%%%%%%%%
%%%%%%%%%%%%%%%%%%%%%%%%%%%%%%%%%%%%%%%%%%%%%%%%%%%%%%%%%%%%%%%%%%%%%
\appendix

% Appendix
%%%%%%%%%%%%%%%%%%%%%%%%%%%%%%%%%%%%%%%%%%%%%%%%%%%%%%%%%%%%%%%%%%%%%
%%%%%%%%%%%%%%%%%%%%%%%%%%%%%%%%%%%%%%%%%%%%%%%%%%%%%%%%%%%%%%%%%%%%%
%%%%%%%%%%%%%%%%%%%%%%%%%%%%%%%%%%%%%%%%%%%%%%%%%%%%%%%%%%%%%%%%%%%%%
\section{Boundary eigenequations}
\label{app:boundaryeigen}
The tight-binding equation of the bulk components of the EBH model is given in (\ref{eq:eigeneqebh}). 
For completeness, we provide below the tight-binding equation satisfied by all components, both bulk and boundaries, that is, for $j,l=1,2,\dots,N$,
\begin{widetext}
	\begin{eqnarray}
		E\psi_{j,l}=&-&J\big[e^{i\phi}(\bar{\delta}_{j,1}\psi_{j-1,l}+\bar{\delta}_{l,1}\psi_{j,l-1}) + e^{-i\phi}(\bar{\delta}_{j,N}\psi_{j+1,l}+\bar{\delta}_{l,N}\psi_{j,l+1})\big]   \nonumber
		\\
		&+& \Delta^\prime \delta_{j,l} \big[\bar{\delta}_{j,1}\psi_{j-1,l-1}+\bar{\delta}_{j,N}\psi_{j+1,l+1}\big] + U\delta_{j,l}\psi_{j,l} + V\delta_{j,l\pm 1}\psi_{j,l}.
		\label{eq:eigeneqebhapp}
	\end{eqnarray}
\end{widetext}
Accordingly, the tight-binding equation of the symmetrized components, when applying (\ref{eq:bosonization}), becomes
\begin{widetext}
	\begin{eqnarray}
		E\psi_{j,l}^B=&-&J\tilde{\delta}_{j,l}\big[e^{i\phi}\bar{\delta}_{j,1}\psi_{j-1,l}^B+e^{-i\phi}\bar{\delta}_{l,N}\psi_{j,l+1}^B+\tilde{\delta}_{i,j-1}\bar{\delta}_{j,l}(e^{i\phi}\bar{\delta}_{l,1}\psi_{j,l-1}^B+e^{-i\phi}\bar{\delta}_{j,N}\psi_{j+1,l}^B)\big]   \nonumber
		\\
		&+& \Delta^\prime\delta_{j,l}\big[\bar{\delta}_{j,1}\psi_{j-1,l-1}^B+\bar{\delta}_{j,N}\psi_{j+1,l+1}^B\big]+U\delta_{j,l}\psi_{j,l}^B + V\delta_{j,l-1}\psi_{j,l}^B,
		\label{eq:eigeneqebhbosapp}
	\end{eqnarray}
\end{widetext}
with $l=1,2.\dots,N$ and $j=1,2,\dots,l$.

% Appendix
%%%%%%%%%%%%%%%%%%%%%%%%%%%%%%%%%%%%%%%%%%%%%%%%%%%%%%%%%%%%%%%%%%%%%
%%%%%%%%%%%%%%%%%%%%%%%%%%%%%%%%%%%%%%%%%%%%%%%%%%%%%%%%%%%%%%%%%%%%%
%%%%%%%%%%%%%%%%%%%%%%%%%%%%%%%%%%%%%%%%%%%%%%%%%%%%%%%%%%%%%%%%%%%%%
\section{Electrical circuit eigenequations}
\label{app:circuiteigen}
Let us consider the $N\times N$ grounded circuit with the design of Fig.~\ref{fig:eleccirc}(a), where $N=7$, whose circuit eigenequation can be written, for the bulk modes defined by $j,l=2,3,\dots,N-1$, as
\begin{widetext}
	\begin{eqnarray}
		\frac{I_{j,l}}{i\omega C}=j(\omega)\mathcal{V}_{j,l}=&\big[&4(1+\delta_{j,l}+\delta_{j,l\pm1})-\frac{1}{\omega^2 LC}(1+\delta_{j,l\pm 1})\big]\mathcal{V}_{j,l}+ \frac{1}{\omega^2 L C}\delta_{j,l\pm 1} \mathcal{V}_{l,j}  \nonumber
		\\
		&-&\big[\mathcal{V}_{j-1,l}+\mathcal{V}_{j+1,l}+\mathcal{V}_{j,l-1}+\mathcal{V}_{j,l+1} + \delta_{j,l}(\mathcal{V}_{j-1,l-1}+\mathcal{V}_{j+1,l+1})\big] .
		\label{eq:eigencircuit}
	\end{eqnarray}
\end{widetext}
Assuming that $\omega=2\pi f$ is the resonant angular frequency of an eigenstate, defined as the frequency for which $j(\omega)=0$ and, therefore, $\mathbf{I}=\mathbf{0}$, the expression in (\ref{eq:eigencircuit}) is simplified to 
\begin{widetext}
	\begin{eqnarray}
		\Big(\frac{f_0^2}{f^2}-4\Big)\mathcal{V}_{j,l}=&-&\big[\mathcal{V}_{j-1,l}+\mathcal{V}_{j+1,l}+\mathcal{V}_{j,l-1}+\mathcal{V}_{j,l+1}\big] - \delta_{j,l}\big[\mathcal{V}_{j-1,l-1}+\mathcal{V}_{j+1,l+1}\big]  \nonumber
		\\
		&+& 4\delta_{j,l}\mathcal{V}_{j,l}+\Big(4-\frac{f_0^2}{f^2}\Big)\delta_{j,l\pm 1}\mathcal{V}_{j,l} + \frac{f_0^2}{f^2}\delta_{i,j\pm 1} \mathcal{V}_{l,j}.
		\label{eq:eigencircuit2}
	\end{eqnarray}
\end{widetext}
where $f_0=\sqrt{\frac{1}{L C}}$ is the natural frequency.

% SUBSECTION
%%%%%%%%%%%%%%%%%%%%%%%%%%%%%%%%%%%%%%%%%%%%%%%%%%%%%%%%%%%%%%%%%%%%%
%%%%%%%%%%%%%%%%%%%%%%%%%%%%%%%%%%%%%%%%%%%%%%%%%%%%%%%%%%%%%%%%%%%%%
%%%%%%%%%%%%%%%%%%%%%%%%%%%%%%%%%%%%%%%%%%%%%%%%%%%%%%%%%%%%%%%%%%%%%
\subsection{Bosonic modes}
With respect to the main diagonal, we define the symmetric ``bosonic'' modes as
\begin{equation}
	\begin{cases}
		\mathcal{V}^B_{j,l}:=\frac{1}{\sqrt{2}}(\mathcal{V}_{j,l}+\mathcal{V}_{l,j}),\ \  \text{for}\  j<l,
		\\
		\mathcal{V}^B_{j,l}:=\mathcal{V}_{j,l},\ \  \text{for}\  j=l,
	\end{cases}
\end{equation}
for which (\ref{eq:eigencircuit2}) can be recast as
\begin{widetext}
	\begin{eqnarray}
		j^\prime(\omega)\mathcal{V}^B_{j,l}=&-&\tilde{\delta}_{j,l}\big[\mathcal{V}^B_{j-1,l}+\mathcal{V}^B_{j,l+1}+\tilde{\delta}_{j,l-1}\bar{\delta}_{j,l}(\mathcal{V}^B_{j,l-1}+\mathcal{V}^B_{j+1,l})\big] \nonumber
		\\
		&-&\delta_{j,l}(\mathcal{V}^B_{j-1,l-1}+\mathcal{V}^B_{j+1,l+1}) + 4\delta_{j,l}\mathcal{V}^B_{j,l}+4\delta_{j,l-1}\mathcal{V}^B_{j,l}, 
		\label{eq:eigencircuitbosapp}
	\end{eqnarray}
\end{widetext}
where $j^\prime(\omega)=\big(\frac{\omega_0^2}{\omega^2}-4\big)=\big(\frac{f_0^2}{f^2}-4\big)$.
Upon applying the gauge transformation defined as
\begin{equation}
	\mathcal{V}^B_{j,l}\to e^{-i \phi(j+l)}\mathcal{V}^B_{j,l},
\end{equation}
we can rewrite (\ref{eq:eigencircuitbosapp}) as
\begin{widetext}
	\begin{eqnarray}
		j^\prime(\omega)\mathcal{V}^B_{j,l}=&-&\tilde{\delta}_{j,l}\big[e^{i\phi}\mathcal{V}^B_{j-1,l}+e^{-i\phi}\mathcal{V}^B_{j,l+1}+\tilde{\delta}_{j,l-1}\bar{\delta}_{j,l}(e^{i\phi}\mathcal{V}^B_{j,l-1}+e^{-i\phi}\mathcal{V}^B_{j+1,l})\big] \nonumber
		\\
		&-&\delta_{j,l}(e^{2i\phi}\mathcal{V}^B_{j-1,l-1}+e^{-2i\phi}\mathcal{V}^B_{j+1,l+1}) + 4\delta_{j,l}\mathcal{V}^B_{j,l}+4\delta_{j,l-1}\mathcal{V}^B_{j,l}, 
		\label{eq:eigencircuitbosapp2}
	\end{eqnarray}
\end{widetext}
which, for $\phi=\frac{\pi}{2}$, reduces to the eigenequation (\ref{eq:eigencircuitbosmain}) of the main text.

% SUBSECTION
%%%%%%%%%%%%%%%%%%%%%%%%%%%%%%%%%%%%%%%%%%%%%%%%%%%%%%%%%%%%%%%%%%%%%
%%%%%%%%%%%%%%%%%%%%%%%%%%%%%%%%%%%%%%%%%%%%%%%%%%%%%%%%%%%%%%%%%%%%%
%%%%%%%%%%%%%%%%%%%%%%%%%%%%%%%%%%%%%%%%%%%%%%%%%%%%%%%%%%%%%%%%%%%%%
\subsection{Fermionic modes}
The eigenequation (\ref{eq:eigencircuit2}) also supports antisymmetric ``(spinless) fermionic" modes defined as
\begin{equation}
	\begin{cases}
		\mathcal{V}^F_{j,l}:=\frac{1}{\sqrt{2}}(\mathcal{V}_{j,l}-\mathcal{V}_{l,j}),\ \  \text{for}\  j<l,
		\\
		\mathcal{V}^F_{j,l}:=0,\ \  \text{for}\  j=l,
	\end{cases}
\end{equation}
whose bulk eigenequation, for $l=3,4,5,6$ and $j=2,3,\dots,l-1$, reads as
\begin{widetext}
	\begin{equation}
		j^\prime(\omega)\mathcal{V}^F_{j,l}=-\big[\mathcal{V}^F_{j-1,l}+\mathcal{V}^F_{j,l+1}+\bar{\delta}_{j,l-1}(\mathcal{V}^F_{j,l-1}+\mathcal{V}^F_{j+1,l})\big]+\Big(4-2\frac{f_0^2}{f^2}\Big)\delta_{j,l-1}\mathcal{V}^F_{j,l}.
		\label{eq:eigencircuitferapp}
	\end{equation}
\end{widetext}
Notice that the $2\frac{f_0^2}{f^2}$ factor at the last term comes from the diagonal inductors connecting the orange nodes in Fig.~\ref{fig:eleccirc}(a), and it is absent in the last term of (\ref{eq:eigencircuitbosapp2}).
In other words, the diagonal inductors only couple to the fermionic modes and not to the bosonic ones.

% SUBSECTION
%%%%%%%%%%%%%%%%%%%%%%%%%%%%%%%%%%%%%%%%%%%%%%%%%%%%%%%%%%%%%%%%%%%%%
%%%%%%%%%%%%%%%%%%%%%%%%%%%%%%%%%%%%%%%%%%%%%%%%%%%%%%%%%%%%%%%%%%%%%
%%%%%%%%%%%%%%%%%%%%%%%%%%%%%%%%%%%%%%%%%%%%%%%%%%%%%%%%%%%%%%%%%%%%%
\subsection{Full eigenequation}

Considering all voltage components $\mathcal{V}_{j,l}$, both bulk and boundaries ones, with $j,l=1,2,\dots,N$, the eigenequation in (\ref{eq:eigencircuit2}) can be generalized as
\begin{widetext}
	\begin{eqnarray}
		\Big(\frac{f_0^2}{f^2}-4\Big)\mathcal{V}_{j,l}=&-&\big[\bar{\delta}_{j,1}\mathcal{V}_{j-1,l}+\bar{\delta}_{j,N}\mathcal{V}_{j+1,l}+\bar{\delta}_{l,1}\mathcal{V}_{j,l-1}+\bar{\delta}_{l,N}\mathcal{V}_{j,l+1}\big] - \delta_{j,l}\big[\bar{\delta}_{j,1}\mathcal{V}_{j-1,l-1}+\bar{\delta}_{j,N}\mathcal{V}_{j+1,l+1}\big]  \nonumber
		\\
		&+& 4\delta_{j,l}\mathcal{V}_{j,l}+\Big(4-\frac{f_0^2}{f^2}\Big)\delta_{i,j\pm 1}\mathcal{V}_{i,j} + \frac{f_0^2}{f^2}\delta_{i,j\pm 1} \mathcal{V}_{l,j}.
	\end{eqnarray}
\end{widetext}
Accordingly, the bosonic eigenequation (\ref{eq:eigencircuitbosapp2}) generalizes as
\begin{widetext}
	\begin{eqnarray}
		j^\prime(\omega)\mathcal{V}^B_{j,l}=&-&\tilde{\delta}_{j,l}\big[e^{i\phi}\bar{\delta}_{j,1}\mathcal{V}^B_{j-1,l}+e^{-i\phi}\bar{\delta}_{l,N}\mathcal{V}^B_{j,l+1}+\tilde{\delta}_{j,l-1}\bar{\delta}_{j,l}(e^{i\phi}\bar{\delta}_{l,1}\mathcal{V}^B_{j,l-1}+e^{-i\phi}\bar{\delta}_{j,N}\mathcal{V}^B_{j+1,l})\big] \nonumber
		\\
		&-&\delta_{j,l}(e^{2i\phi}\bar{\delta}_{j,1}\mathcal{V}^B_{j-1,l-1}+e^{-2i\phi}\bar{\delta}_{j,N}\mathcal{V}^B_{j+1,l+1}) + 4\delta_{j,l}\mathcal{V}^B_{j,l}+4\delta_{j,l-1}\mathcal{V}^B_{j,l}, 
	\end{eqnarray}
\end{widetext}
with $l=1,2,\dots,N$ and $j=1,2,\dots,l$, while the fermionic eigenequation (\ref{eq:eigencircuitferapp}) generalizes as
\begin{widetext}
	\begin{equation}
		j^\prime(\omega)\mathcal{V}^F_{j,l}=-\big[\bar{\delta}_{j,1}\mathcal{V}^F_{j-1,l}+\bar{\delta}_{l,N}\mathcal{V}^F_{j,l+1}+\bar{\delta}_{l,1}\bar{\delta}_{j,l-1}(\mathcal{V}^F_{j,l-1}+\bar{\delta}_{j,N}\mathcal{V}^F_{j+1,l})\big]+\Big(4-2\frac{f_0^2}{f^2}\Big)\delta_{j,l-1}\mathcal{V}^F_{j,l},
	\end{equation}
\end{widetext}
with $l=1,2,\dots,N$ and $j=1,2,\dots,l-1$.

% SECTION
%%%%%%%%%%%%%%%%%%%%%%%%%%%%%%%%%%%%%%%%%%%%%%%%%%%%%%%%%%%%%%%%%%%%%
%%%%%%%%%%%%%%%%%%%%%%%%%%%%%%%%%%%%%%%%%%%%%%%%%%%%%%%%%%%%%%%%%%%%%
%%%%%%%%%%%%%%%%%%%%%%%%%%%%%%%%%%%%%%%%%%%%%%%%%%%%%%%%%%%%%%%%%%%%%
\section*{Acknowledgments}
\label{sec:acknowledgments}

This work was developed within the scope of the Portuguese Institute for Nanostructures, Nanomodelling, and Nanofabrication (i3N) Projects No. UIDB/50025/2020, No. UIDP/50025/2020, and No. LA/P/0037/2020, financed by national funds through the Funda\c{c}\~ao para a Ci\^encia e Tecnologia (FCT) and the Minist\'erio da Educa\c{c}\~ao e Ci\^encia
(MEC) of Portugal. 
The author acknowledges financial support from i3N through the work Contract No. CDL-CTTRI-46-SGRH/2022, and from FCT through the work Contract No.~CDL-CTTRI-91-SGRH/2024. The author would also like to thank Ricardo Dias for many insightful comments and suggestions in the final stages of preparation of the manuscript, and also Rui Martins for fruitful discussions.

\bibliography{majoranabiblio}
%\cleardoublepage

\end{document}